\documentclass[twocolumn]{aastex631}
\usepackage{showyourwork}

\newcommand{\sbu}{Department of Physics and Astronomy, Stony Brook University, Stony Brook NY 11794, USA}
\newcommand{\cca}{Center for Computational Astrophysics, Flatiron Institute, New York NY 10010, USA}

\usepackage{bbold}

\usepackage[commandnameprefix=always,commentmarkup=uwave]{changes}
\definechangesauthor[name={Sabina}, color=cyan]{S}
\definechangesauthor[name={Will}, color=magenta]{W}
\definechangesauthor[name={Brett}, color=yellow]{B}
\definechangesauthor[name={Rodrigo}, color=green]{R}

\DeclareMathOperator{\diag}{diag}

\usepackage{hyperref}  
\usepackage{fontawesome5}  

\begin{document}

\title{Polka-dotted Stars: a Hierarchical Model for Mapping Stellar Surfaces Using Occultation Light Curves and the Case of TOI-3884}

\author[0000-0002-6650-3829]{Sabina Sagynbayeva}
\email{sabina.sagynbayeva@stonybrook.edu}
\affiliation{\sbu}
\affiliation{\cca}
\affiliation{Kavli Institute for Theoretical Physics, University of California, Santa Barbara, CA 93106, USA}

\author[0000-0003-1540-8562]{Will M. Farr}
\affiliation{\sbu}
\affiliation{\cca}

\author[0000-0003-2528-3409]{Brett M. Morris}
\affiliation{Space Telescope Science Institute, 3700 San Martin Dr, Baltimore, MD 21218, USA}

\author[0000-0002-0296-3826]{Rodrigo Luger}
\affiliation{\cca}

\begin{abstract}
    We present \texttt{StarryStarryProcess}, a novel hierarchical Bayesian
    framework for mapping stellar surfaces using exoplanet transit light curves.
    While previous methods relied solely on stellar rotational light curves -—
    which contain limited information about spot properties -- our approach
    leverages planetary transits as probes of stellar surfaces. When a planet
    crosses a spot during transit, it creates a distinctive
    change in the light curve that directly reveals
    spot properties. Our model integrates planetary transit modeling with
    stellar variability analysis by combining the spherical harmonic
    surface map representation from \texttt{starry}, the
    probabilistic approach to spot properties of
    \texttt{StarryProcess}, and a comprehensive transit model that accounts for
    spot-crossing events during transits. We demonstrate through
    synthetic data experiments that our model successfully recovers spot
    distributions, stellar orientation, and spot physical properties. We extend
    the framework to handle evolving stellar surfaces through time-dependent
    modeling. Applying our method to TESS observations of TOI-3884, we find
    evidence for high-latitude spot concentrations and significant spin-orbit
    misalignment. The transit-based approach overcomes fundamental limitations
    of previous models by providing constraints on spot properties that would
    remain hidden in the null space of rotational light curves alone. This
    methodology enables more accurate exoplanet characterization by
    disentangling stellar activity due to starspots from planetary signals while simultaneously
    providing insights into stellar magnetic activity patterns. The whole paper is reproducible, 
    and can be found by clicking the GitHub icon \href{https://github.com/ssagynbayeva/polka-dotted-stars-toi3884/tree/main}{\faGithub}.
        
\end{abstract}

\section{Introduction}
\label{sec:intro}
The detection of atmospheres on Earth analogs using transmission spectroscopy is challenging due to stellar 
contamination effects \citep{Ducrot2018, Morris2018}. Stellar magnetic activity, particularly starspots, significantly 
influences observed transmission spectra, making it difficult to distinguish planetary atmospheric signals from stellar ones. 
Starspots exhibit variations in both time and wavelength primarily due to stellar rotation, potentially complicating the analysis 
and interpretation of exoplanet observations.

Therefore, characterizing stellar surface features arising from magnetic activity—-such as starspots and flares—-is crucial for 
improving planet detection efficiency across photometric, spectral, and radial velocity studies. Understanding starspots is 
especially important for the JWST mission, which aims to characterize numerous exoplanet atmospheres around stars with strong 
variability. Recent JWST results on TRAPPIST-1b \citep{Lim2023} demonstrated that accurate measurements have been 
hindered by stellar contamination, highlighting the need for better models to account for these effects. To fully leverage 
JWST's capabilities and accurately interpret transmission spectra, we must develop a comprehensive understanding of stellar 
magnetic activity and its impact on observed signals.

Recent works have provided valuable insights into the topographical
characteristics of stars and their magnetic activity. For example, starspot
influence can be observed through photometric data from Kepler or TESS. The
\texttt{starry} code \citep{Luger2019} offers an efficient method for computing
light curves of stars and planets using a spherical harmonic framework to
represent each object's surface brightness distribution,
enabling fast and accurate calculations of rotational light curves and
occultations. Building upon \texttt{starry}, \citet{Luger2021b} introduced
\texttt{StarryProcess}, a probabilistic model for stellar variability that
uses a model for the spot distribution
to derive a Gaussian process (GP) for the
representation of the surface brightness in spherical harmonic space, allowing
efficient modeling of stellar variability over extended time scales.

With \texttt{StarryProcess}, \citet{Luger2021b} investigated stellar spot
features using Gaussian Processes with a physically interpretable kernel. They
argued that complex degeneracies make it impossible to determine the precise
configuration of starspots solely from rotational light curves. Instead, they
suggested focusing on the statistical properties of spot distributions rather
than individual spot characteristics. However, they also demonstrated that
constraining starspot properties from a single star's light curve is difficult;
an ensemble analysis of multiple stars is more informative.

While previous models like \texttt{StarryProcess} relied solely on stellar
rotational light curves---which contain limited information about spot
properties \citep{Luger2021a}---our approach introduces a critical advancement:
we leverage planetary transits as probes of stellar surfaces while also simultaniously fitting the full light curve. 
Exoplanet host
stars are \textit{uniquely suited} for photometric studies of starspot
properties because: (1) Planetary transits effectively scan different regions of
the stellar disk; (2) Transit light curves contain encoded information about the
stellar surface and starspots; (3) Spot-crossing events during transits provide
direct constraints on spot brightness, size, and latitude.

\begin{figure}[hbt!]
    \includegraphics[width=0.46\textwidth]{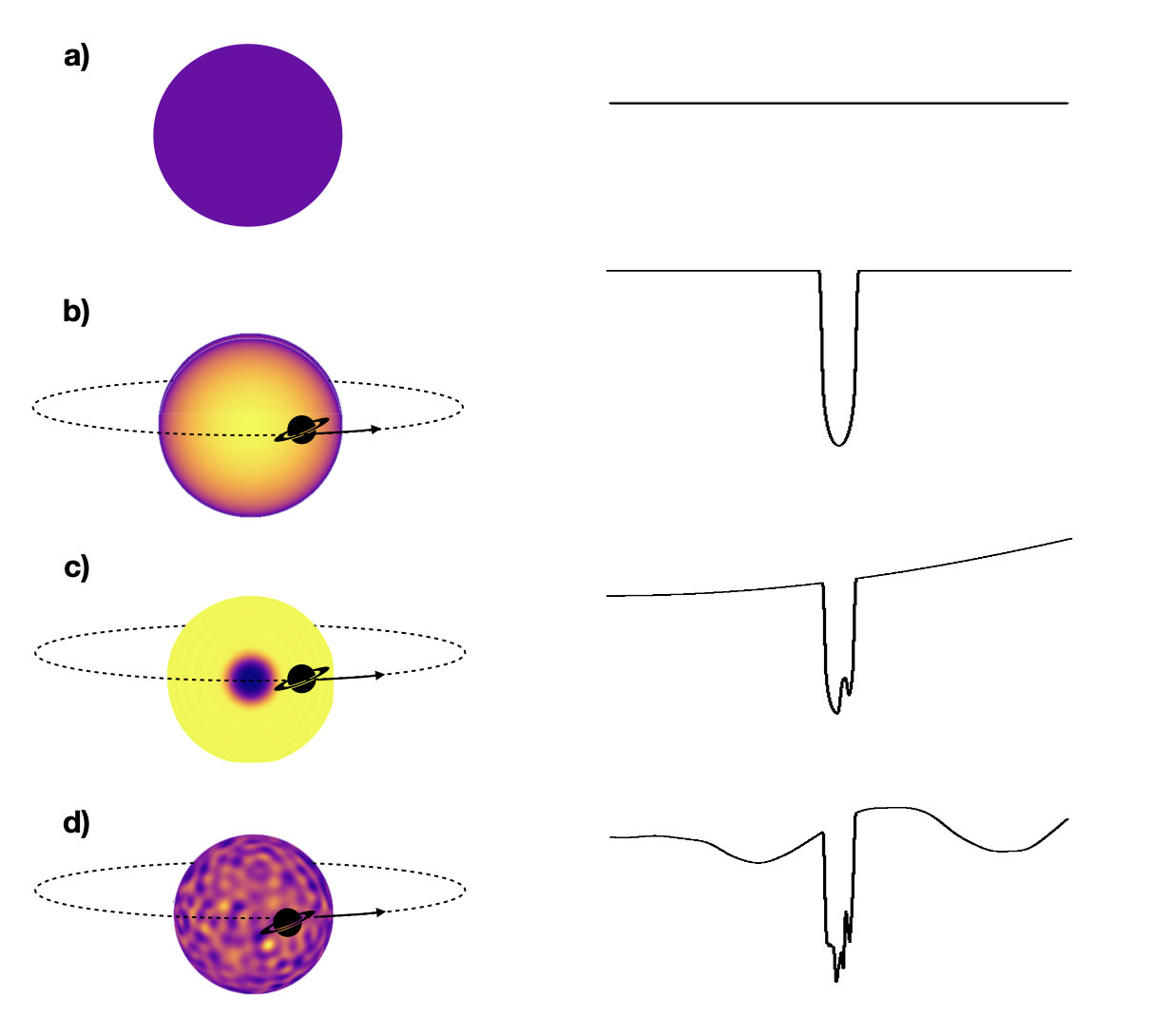}
        \caption{Light curves from the kinds of stars shown on the left panel (these light curves were calculated analytically 
        using \texttt{starry} \citep{Luger2019}): (a) the light curve shows a flat flux if the star has no surface features and no planet orbiting it; 
        (b) a transit in the light curve of the star that hosts an exoplanet; (c) if the star has one orbiting planet and one starspot under the planet's 
        trajectory then it is seen as a bump in a transit signal; (d) the light curve becomes complicated when the star has multiple spots.
        In this illustration, the planet has rings to make it distinct from a dark circular spot.}
        \label{fig:cartoon}
    \end{figure}

This transit-based approach overcomes the fundamental limitation of previous models that analyzed stellar light curves alone, 
where too many degeneracies exist to precisely characterize spots. As illustrated in Figure \ref{fig:cartoon}, when a planet 
crosses a spot during transit, it creates a distinctive bump in the light curve that directly reveals spot properties. Transit mapping approach was 
first introduced by \cite{Silva2003}. This foundational work sparked extensive development of models that account for 
spot characteristics, including 
\cite{Huber2010,Oshagh2013,Tregloan-Reed2013,Tregloan-Reed2015,Beky2014,Montalto2014,Maxted2016,Herrero2016,Juvan2018,Morris2017,Scandariato2017}.

In this work, we develop \texttt{StarryStarryProcess}, a novel framework that integrates planetary transit modeling with 
stellar variability analysis. Our model combines:
\begin{itemize}
    \item The spherical harmonic representation from \texttt{starry}
    \item The probabilistic approach of \texttt{StarryProcess}
    \item A comprehensive transit model that accounts for spot-crossing events
    \item The capability to model evolving stellar surfaces
    through efficient interpolation of surface maps
\end{itemize}

By simultaneously modeling stellar variability, spot distributions, and
planetary transits, our approach disentangles spot effects from transit
features, resulting in more accurate constraints on both stellar and planetary
parameters. When modeling multiple transits alongside underlying stellar
variability, we can better determine stellar inclination, obliquity, and spot
distribution patterns as well as providing constraints on the
spot patterns in the surface map that lie beneath the transit.

\begin{figure*}[hbt!]
    \includegraphics[width=\textwidth]{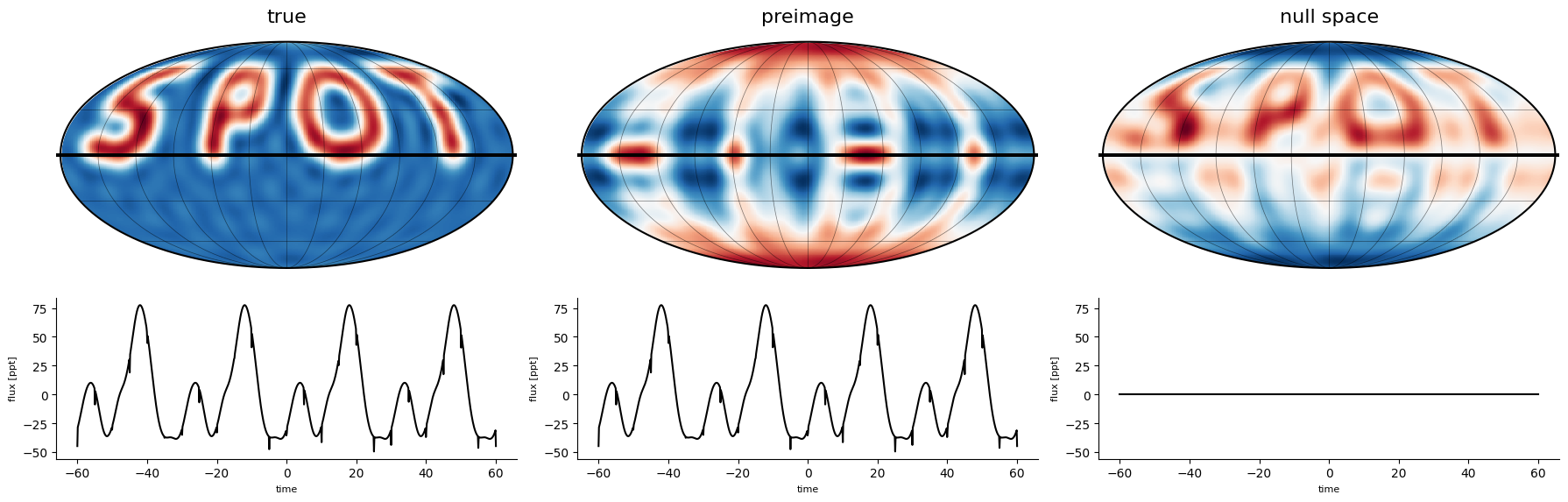}
        \caption{The breakdown of a surface map into its constituent components: the original map (left), its preimage (center), and null space (right). 
        Accompanying each set is the corresponding impact on the star's rotational light curve. The preimage represents surface patterns that 
        directly influence the observed light curve, while the null space encompasses patterns that have no effect on the flux measurements. 
        It's noteworthy that a significant portion of the surface features fall within the null space, meaning they do not contribute to the light curve 
        that we can observe and measure. The black line shows the planetary trajectory, so what is important is that the preimage contains the full
        information under planetary trajectory, while the nullspace is missing it. 
        \href{https://github.com/ssagynbayeva/polka-dotted-stars-toi3884/blob/main/src/tex/notebooks/nullspace-preimage.ipynb}{\faGithub}
        }
        \label{fig:nullspace}
    \end{figure*}

Figure \ref{fig:nullspace} illustrates a key insight of our approach: while
previous methods struggled with the null space problem (where many surface
features don't affect the observable light curve), planetary transits
effectively ``carve out'' paths on the stellar surface that reveal information
otherwise hidden in the null space. The transit trajectory covers regions in the
preimage component (observable) that would remain undetectable
without the presence of a transit.

Our paper is organized as follows: Section \ref{sec:model} presents the
\texttt{StarryStarryProcess} model; Section \ref{sec:results-synthetic} presents all results with 
Section \ref{sec:experiment1} presenting results on non-evolving stellar
surfaces and Section \ref{sec:experiment2} presenting results on evolving surfaces;
Section \ref{sec:toi3884} presents results on TESS data for TOI-3884; and
finally, we discuss implications and assumptions in Section
\ref{sec:discussion}.

\section{Hierarchical Bayesian model}
\label{sec:model}

In this section we describe our probabilistic model for stellar surfaces and the
parameters we use to infer their physical properties.  Our parameters break down
into three groups: we have parameters that describe the \citet{Luger2021b}
Gaussian process from which the spherical harmonic components of the stellar
surface map are drawn; parameters of the star's orientation and size; and the
planet's orbital parameters:
\begin{linenomath}\begin{align}
    \label{eq:largetheta}
    \pmb{\Theta}
     & =
    \left(
    \theta_\bullet
    \,\,\,
    \theta_\star
    \,\,\,
    \theta_p
    \right)^\top
    \quad,
\end{align}\end{linenomath}

The GP parameters describing the distribution of the spherical harmonic
coefficients of the stellar surface are 
\citep{Luger2021b}:
\begin{linenomath}\begin{align}
        \label{eq:thetaspot}
        \pmb{\theta}_\bullet
         & =
        \left(
        \mathbb{n}
        \,\,\,
        \mathbb{c}
        \,\,\,
        \sigma^2_\phi
        \,\,\,
        \mu_\phi
        \,\,\,
        \mathbb{r}
        \right)^\top
        \quad,
    \end{align}\end{linenomath}
where $\mathbb{n}$ is the number of starspots, $\mathbb{c}$ is their contrast
(defined as the intensity difference between the spot and the background
intensity, as a fraction of the background intensity), and $\mathbb{r}$ is the
radius of the spots, $\mu_\phi$ and $\sigma^2_\phi$ are the mean and variance of
(Beta) distribution over latitudes. 

In reality, spot contrast exhibits 
strong wavelength dependence due to the temperature difference between spotted and unspotted photospheric 
regions, with cooler spots appearing more prominent at shorter wavelengths. This is not apparent in a single-band photometry, but would be apparent in multi-band photometry. Though we do not do it here, it would be straightforward to impose a temperature-dependence on contrast parameter, and use a multi-band photometry to infer the temperature rather than the phenomenological contrast parameter as we do it here  \citep[e.g.,][]{Herrero2016,Juvan2018}. Moreover, because our analysis assumes that the changes in brightness from spots are small, we are unable to distiguish between bright and dark spots on stellar surfaces. An analysis that treated the effect of spots in full non-linearity might be able to distiguish these but would lose the computational efficiency of marginalization that we exploit.
We should remember that a model using discrete, circular spots with uniform contrast is probably equally unrealistic than our approach. Our model offers greater adaptability compared to discrete spot models, making it potentially more effective for representing actual stellar surfaces.
Our current parameterization was adopted to maintain computational tractability while establishing the 
foundational framework. 
\cite{Morris2018} analyzed simultaneous light curves from both Kepler and Spitzer observations of TRAPPIST-1, concluding that a bright spot model better explains the stellar behavior compared to dark spot models. A comprehensive analysis of how multi-wavelength photometry affects the ambiguities inherent in stellar mapping techniques is outside the scope of this current paper.

\citet{Luger2021b} suggests that it is more convenient to sample $\mu_\phi$ and
$\sigma^2_\phi$ using $\mathbb{a}$ and $\mathbb{b}$, where $\mathbb{a}$ and
$\mathbb{b}$ are the normalized parameters of the Beta distribution in
$\cos\phi$, which is the probability density function (PDF) for the latitude
$\phi$ of the spots. These parameters have a one-to-one correspondence to the
mode $\mu_\phi$ and standard deviation $\sigma_\phi$ of the distribution in
$\phi$, allowing for a concise representation of the latitudinal distribution
characteristics. The Beta distribution in $\cos\phi$ which has hyperparameters
$\alpha$ and $\beta$, and the PDF given by
\begin{align}
    \label{eq:cosphi-pdf}
    p \big(\cos\phi \, \big| \, \alpha, \beta \big)
     & =
    \dfrac{\Gamma(\alpha + \beta)}{\Gamma(\alpha)\Gamma(\beta)}
    (\cos\phi)^{\alpha - 1}
    (1 - \cos\phi)^{\beta - 1}
    \quad,
\end{align}
where $\Gamma$ is the Gamma function. The $\alpha$ and $\beta$ are derived from the normalized parameters 
\begin{align}
    \label{eq:beta2gauss}
    \alpha & = \exp\left({(\ln\frac{1}{2})\mathbb{a}}\right)
    \nonumber                                                 \\
    \beta  & = \exp\left({\ln\frac{1}{2} + (10 - \ln\frac{1}{2})\mathbb{b}}\right)
    \quad,
\end{align}
with inverse transform
\begin{align}
    \label{eq:gauss2beta}
    \mathbb{a} & \equiv \frac{\ln\alpha}{\ln\frac{1}{2}}
    \nonumber                                             \\[0.5em]
    \mathbb{b} & \equiv \frac{\ln\beta - \ln\frac{1}{2}}{10 -\ln\frac{1}{2}}
\end{align}
The parameters $\mathbb{a}$ and $\mathbb{b}$ are both constrained to values between 0 and 1, making them convenient for sampling during the inference process. 
However, $\mathbb{a}$ and $\mathbb{b}$ do not have a straightforward relationship with physically meaningful quantities. In many situations, it is preferable to parametrize 
the latitude distribution using two parameters: $\mu_\phi$, which controls the central latitude, and $\sigma_\phi$, which governs the dispersion of the 
spots' latitudes. Here, both the mean $\mu_\phi$ and variance $\sigma^2_\phi$ can be derived from the Beta distribution as
\begin{align}
    \label{eq:mean_beta}
    \mu_\phi
     & =
    \dfrac{\alpha}{\alpha+\beta}
    \quad,
\end{align}
\begin{align}
    \label{eq:var_beta}
    \sigma^2_\phi
     & =
    \dfrac{\alpha\beta}{(\alpha+\beta)^2(\alpha+\beta+1)}
    \quad,
\end{align}
Following \citet{Luger2021b}, these parameters are used to define a mean and
covariance matrix for the Gaussian distribution of the spherical harmonic
surface map components of the star.  

The stellar parameters are 
\begin{linenomath}\begin{align}
    \label{eq:thetastar}
    \pmb{\theta_\star}
     & =
    \left(
    i_\star
    \,\,\,
    m_\star
    \,\,\,
    u_1
    \,\,\,
    u_2
    \,\,\,
    P_\star
    \right)^\top
    \quad,
\end{align}\end{linenomath}
where $i_\star$ is the star's orbital inclination, $m_\star$ is the stellar mass in the units of the solar mass, $u_1$ and $u_2$ are limb-darkening coefficients,
and $P_\star$ is the rotational period of the star.

The planetary parameters are 
\begin{linenomath}\begin{align}
    \label{eq:thetap}
    \pmb{\theta_p}
     & =
    \left(
    i_p
    \,\,\,
    e
    \,\,\,
    \psi
    \,\,\,
    \omega
    \,\,\, P \,\,\, t_0 \,\,\, R_p/R_\star \right)^\top \quad,
    \end{align}
\end{linenomath} 
where $i_p$ is the planet's orbital inclination, $e$ is its eccenticity,
$\psi$ is the stellar obliquity (beacuse the lightcurve is
invariant under rotations of the star-planet system in the plane of the sky,
only the \emph{relative} obliquity between the planet's orbit and the star's
spin matter for the lightcurve; we, by convention, ignore the stellar obliquity,
and sample over the planet's), $\omega$ is the argument of pericenter of the
planet, $P$ is the orbital period of the planet, $t_0$ is the transit start
time, and $R_p/R_\star$ is the planet to star radius ratio.

The \texttt{starry} model of \citet{Luger2019} writes the flux
$\mathbb{f}_\mathrm{true}$ from the system at the observation times as a linear
function of the spherical harmonic surface maps $\mathbb{y}$, using a design
matrix $\pmb{M}\left( \pmb{\Theta} \right)$ that depends on the orbital and
rotational properties of the system:
\begin{equation}
    \mathbb{f}_\mathrm{true} = \pmb{M}\left( \pmb{\Theta} \right) \mathbb{y}
\end{equation}
Since our prior for the surface map components $\mathbb{y}$ is Gaussian, and the
\texttt{starry} map from surface map to flux is linear, our model induces a
Gaussian prior distribution for the true flux, $\mathbb{f}_{true}$, with
constant mean and a covariance matrix that can be expressed analytically in
terms of the design matrix \citep{Luger2021b}. 

We assume that the flux observations $\mathbb{f}_{obs}$ are contaminated by
additive Gaussian noise, such that:
\begin{equation}
    \mathbb{f}_\mathrm{obs} = \mathbb{f}_\mathrm{true} + \epsilon
\end{equation}
where $\epsilon \sim \mathcal{N}\left(0, \mathrm{diag}\left( \vec{\sigma}
\right)\right)$ is the noise term with independent Gaussian noise with standard
deviations given by $\vec{\sigma}$ at each observation. 

Given the GP prior and the likelihood function, we can calculate the joint
posterior distribution over the hyperparameters $\pmb{\Theta}$ and the map
$\mathbb{y}$ given the observed fluxes $\mathbb{f}_{obs}$:
\begin{equation}
    p\left(\pmb{\Theta}, \mathbb{y} \mid \mathbb{f}_{obs}\right) \propto p\left(\pmb{\Theta}\right) p\left(\mathbb{y} \mid \pmb{\Theta}\right) p\left(\mathbb{f}_{obs} \mid \mathbb{y}\right)
\end{equation}
where $p\left(\pmb{\Theta}\right)$ is the prior distribution over the
hyperparameters, $P\left(\mathbb{f}_{true} \mid \pmb{\Theta}\right)$ is the
likelihood of the true flux given the hyperparameters, and $P(\mathbb{f}_{obs}
\mid \mathbb{f}_{true})$ is the likelihood of the observed data given the true
function.  Because the last two terms in this equation are Gaussians, we can
factor them to isolate a Gaussian term containing the map $\mathbb{y}$ and
another Gaussian containing only the parameters $\pmb{\Theta}$ \citep[see,
e.g.,][]{Hogg2020}.  Marginalizing over the map $\mathbb{y}$ eliminates that
term, leaving 
\begin{multline}
    \label{eq:log-likeRodrigo}
    p\left( \mathbb{f}_\mathrm{obs} \mid \pmb{\Theta} \right) = \int \mathrm{d} \mathbb{y} \, p\left( \mathbb{f}_\mathrm{obs} \mid \mathbb{y} \right) p\left( \mathbb{y} \mid \pmb{\Theta} \right) \\ \propto \frac{1}{\sqrt{\left| \pmb{B} \right|}} \exp\left( -\frac{1}{2} \mathbf{r}^T \pmb{B}^{-1} \mathbf{r} \right),
\end{multline}
where $\mathbf{r} = \mathbb{f}_{obs} - \pmb{\mu}\left(\pmb{\Theta} \right)$ is
the residual vector, and
\begin{equation}
    \pmb{B} = \pmb{C} + \pmb{M}\left(\pmb{\Theta}\right) \pmb{\Lambda}\left(\pmb{\Theta} \right) \pmb{M}^T\left(\pmb{\Theta}\right),
\end{equation}
is the marginal data covariance. The marginal data covariance matrix $\pmb{B}$ encodes the total expected variability in our observations 
and has a clear physical interpretation. The first term, $\pmb{C} = \diag\left({\vec{\sigma}^2}\right)$, 
represents the measurement uncertainties—the noise we expect from our instruments. The second term, 
$\pmb{M}\left(\pmb{\Theta}\right) \pmb{\Lambda}\left(\pmb{\Theta} \right) \pmb{M}^T\left(\pmb{\Theta}\right)$, 
captures the intrinsic variability we expect from the stellar surface features themselves. Here, $\pmb{\Lambda}$ is the 
covariance matrix of the spherical harmonic coefficients from our GP model, which describes how different map components 
(corresponding to different spatial scales and patterns on the stellar surface) are expected to vary and correlate with each 
other. The design matrix $\pmb{M}$ translates these surface variations into observable flux variations in our light curve. 
Together, $\pmb{B}$ tells us not only how much each flux measurement should vary, but also how measurements at different times 
should be correlated—for instance, flux variations separated by short time intervals will be more strongly correlated than 
those separated by long intervals, reflecting the smooth evolution of surface features as the star rotates.

The posterior on the hyperparameters $\pmb{\Theta}$ is then
given by
\begin{equation}
    p\left(\pmb{\Theta} \mid \mathbb{f}_{obs}\right) \propto p\left(\pmb{\Theta}\right) p\left( \mathbb{f}_\mathrm{obs} \mid \pmb{\Theta} \right). 
\end{equation}
We have marginalized the map $\mathbb{y}$ out of the posterior, but the
conditional distribution of $\mathbb{y}$ given the data at fixed parameters
$\pmb{\Theta}$ is Gaussian, and so we can draw samples from it
straightforwardly; formulas are given in \citet{Hogg2020}.  Also as discussed in
\citet{Hogg2020}, the inverse of $\pmb{B}$ can be computed much more efficiently than the naive expectation%
\footnote{$\pmb{B}$ has dimensions $K \times K$, where $K$ is the number of data points, and the naive expectation would be $K^3$ operations} %
using the Sherman-Morrison-Woodbury matrix identity:
\begin{multline}
    \label{eq:Hoggtrick}
    \pmb{B}^{-1} = \\ \pmb{C}^{-1} - \pmb{C}^{-1} \pmb{M} \left( \pmb{\Lambda}^{-1} + \pmb{M}^T \pmb{C}^{-1} \pmb{M} \right)^{-1} \pmb{M}^T \pmb{C}^{-1};
\end{multline}
this requires only inverting the diagonal matrix $\pmb{C}$ and matrices that are
square with the size of the number of map components.


\subsection{The issue of the normalization of light curves and units}
Before discussing the experiments produced for this paper, we first need to
remind the reader of a subtlety of the flux normalization when we are given the
raw light curves from telescopes.  The problem and the ways to tackle it were
described in \cite{Luger2021a} and \cite{Luger2021b}. 

Briefly, the problem of normalization of light curves is related to the fact
that the observed flux from a star can vary due to a variety of factors such as
atmospheric effects, instrumental noise, and changes in the intrinsic brightness
of the star itself. These variations can make it difficult to compare light
curves of different stars or even the same star observed at different times. To
address this problem, astronomers typically normalize light curves by dividing
the observed flux by some factor (the median or mean of the flux) that is
assumed to be constant over time. However, as was described in
\cite{Luger2021a}, if a star has a single large equatorial spot of contrast $c$
viewed at some high inclination (\cite{Luger2021a} used the value of
$60^\circ$), and another star with a spot at the same location but with half the
contrast \emph{and} a large polar spot of comparable contrast, then the light
curves for both of the stars become indistiguishable in the relative units
astronomers observe them. In addition to normalizing the flux, \cite{Luger2021a}
also added a \emph{baseline} (1 in their case), which is the flux one would have
gotten from a spotless star, and is an additive component to the $Y_m^l$.

We do not adopt this approach, however; instead we fit for the constant
component of the map (i.e.\ the $\ell = m = 0$ component of $\mathbb{y}$).  We
impose a prior distribution over these constant terms that is completely
uniform, allowing the data to constrain the baseline.  

\subsection{Stellar inclination and obliquity}
\label{sec:obl-inc}
\begin{figure*}[hbt!]
    \begin{centering}
        \includegraphics[width=\linewidth]{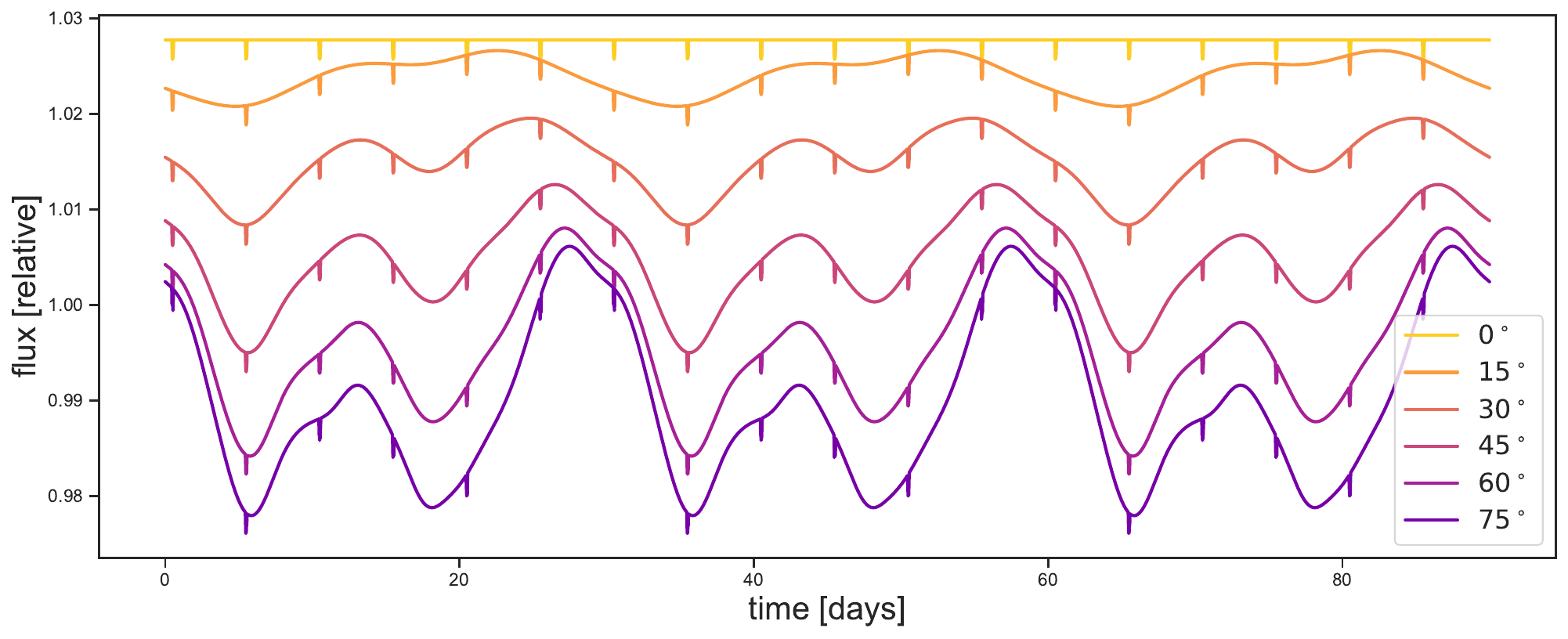}
        \caption{
            Variation in stellar light curves as a function of inclination angle. Each curve displays the photometric time series (light curve) for a 
            hypothetical spotted star observed at different inclination angles. The x-axis represents time in days, while the y-axis shows the relative flux. 
            Inclination angles range from $0^\circ$ (pole-on view) to $90^\circ$ (equator-on view). Note how the amplitude and shape of the light 
            curve modulations change significantly with inclination, demonstrating the importance of this parameter in interpreting stellar variability. 
            The star's rotation period and spot configuration remain constant across all light curves.
            \href{https://github.com/ssagynbayeva/polka-dotted-stars-toi3884/blob/main/src/tex/notebooks/inclination-obliquity.ipynb}{\faGithub}
        }
        \label{fig:inclination}
    \end{centering}
\end{figure*}

In the Section \ref{sec:results-synthetic}, we demonstrate
that there is a good amount of information
about stars that we can learn from disk-integrated photometric measurements
in the presence of transits. In particular, Figure
\ref{fig:inclination} illustrates the profound impact of stellar inclination on
observed light curves. As the inclination angle varies from pole-on ($0^\circ$)
to equator-on ($90^\circ$) views, we observe significant changes in both the
amplitude and morphology of the light curve modulations. At low inclinations,
when we view the star more pole-on, the light curves exhibit smaller amplitude
variations. This is because the spots remain invisible or
visible throughout the stellar rotation, leading to more subtle flux changes.
As the inclination increases towards an equator-on perspective, the amplitude of
the variations typically increases, and the shape changes of the
light curve become more pronounced. These changes occur
because at higher inclinations, spots periodically disappear from view as the
star rotates, causing more dramatic swings in the observed flux. The effect is
particularly noticeable for spots at mid-latitudes, which can produce sharp dips
in the light curve as they rotate into and out of view. It's important to note
that while inclination significantly affects the observed light curve, it does
not alter the underlying rotation period of the star or the intrinsic properties
of the spots themselves.

\begin{figure*}[hbt!]
    \begin{centering}
        \includegraphics[width=\linewidth]{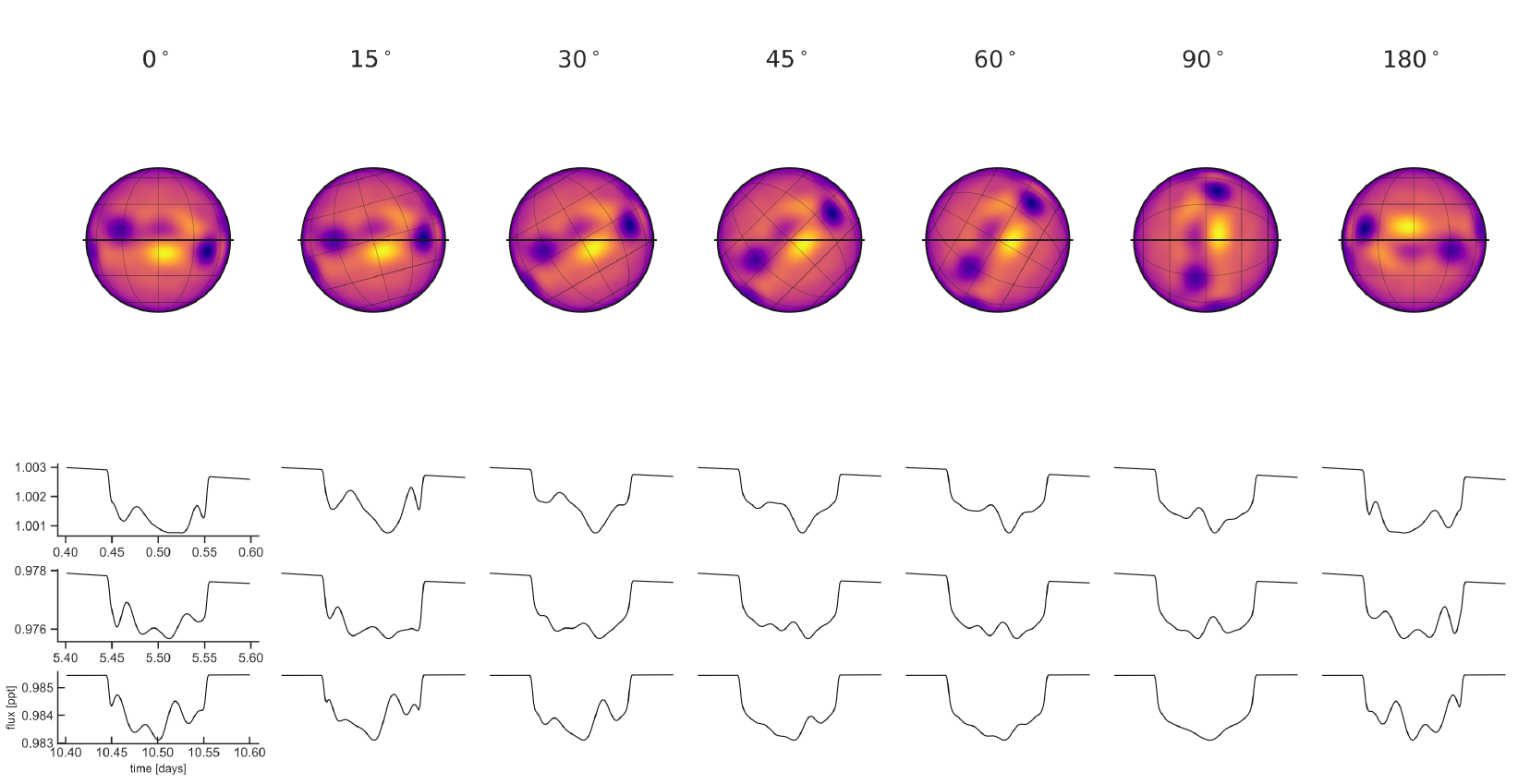}
        \caption{
            Impact of stellar obliquity on consecutive planetary transit light curves. Each column represents a different stellar obliquity scenario, 
            with obliquity increasing from left to right. The rows show three consecutive transits for each scenario. The x-axis represents time relative to 
            the transit center, while the y-axis shows the normalized flux. Note the varying patterns of in-transit brightness fluctuations 
            (spot-crossing events) across different obliquities and between consecutive transits. These differences arise from the changing 
            geometry between the planet's transit path and the distribution of starspots at different stellar latitudes. 
            The consistent transit depth and duration across all panels indicate that other system parameters remain constant, 
            isolating the effect of stellar obliquity.
            \href{https://github.com/ssagynbayeva/polka-dotted-stars-toi3884/blob/main/src/tex/notebooks/inclination-obliquity.ipynb}{\faGithub}
        }
        \label{fig:obliquity}
    \end{centering}
\end{figure*}

While variations in stellar obliquity can have significant implications for planetary systems, their effects have no effect on the rotational modulation. The global light curve is primarily influenced by the star's rotation and the distribution of spots on its surface, 
with obliquity playing a less prominent role. However, the impact of stellar obliquity becomes much more apparent during planetary transits, 
particularly through the phenomenon of spot-crossing events. When a planet transits its host star, it can occult starspots along its path. 
These spot-crossing events manifest as brief increases in brightness during the transit, as the planet temporarily blocks a darker region of the stellar 
surface. The frequency, timing, and duration of these spot-crossing events are strongly influenced by the alignment between the stellar spin axis and 
the planet's orbital plane - in other words, the stellar obliquity. Figure \ref{fig:obliquity} shows three consecutive transit light curves for 
different stellar obliquities. Each row in the figure represents a different obliquity scenario, with obliquity increasing from left to right. 
The rows display three consecutive transits for each scenario. 
In systems with low obliquity, where the stellar equator is aligned with the 
planet's orbit, spot-crossing events tend to occur more regularly and predictably. This is evident in the first column of Figure \ref{fig:obliquity}, where the spot-crossing patterns show a more consistent appearance across consecutive transits. 
Conversely, in systems with high obliquity, the pattern of spot-crossings can be more erratic and less frequent. This is because the planet's transit path across 
the stellar disk intersects a different range of stellar latitudes, potentially missing spots concentrated at certain latitudes. The $90^\circ$ column of 
Figure \ref{fig:obliquity} demonstrates this, showing more varied and less predictable spot-crossing patterns between consecutive transits. 
Therefore, while the full-disk light curve may not clearly reveal changes in stellar obliquity, careful analysis of in-transit light curves, 
particularly focusing on spot-crossing events, can provide valuable insights into the star's orientation relative to the planetary orbit. 

While the analysis of spot-crossing events in transit light curves provides a
powerful tool for constraining stellar inclination and obliquity, there exists
an important degeneracy in these measurements.
A system with stellar inclination $i_\star$
and obliquity $-\psi_\star$ will produce identical light curves to a system
with inclination $180-i_\star$ and obliquity $\psi_\star$, but with a surface
map reflected about the equator.

This degeneracy arises from the symmetry in the geometry of the star-planet system. In both scenarios, the relative orientation between the stellar 
rotation axis and the planet's orbital plane remains the same, merely mirrored. As a result, the pattern of spot crossings and their evolution over 
multiple transits will be indistinguishable between these two configurations.

\begin{figure}[hbt!]
    \begin{centering}
        \includegraphics[width=\linewidth]{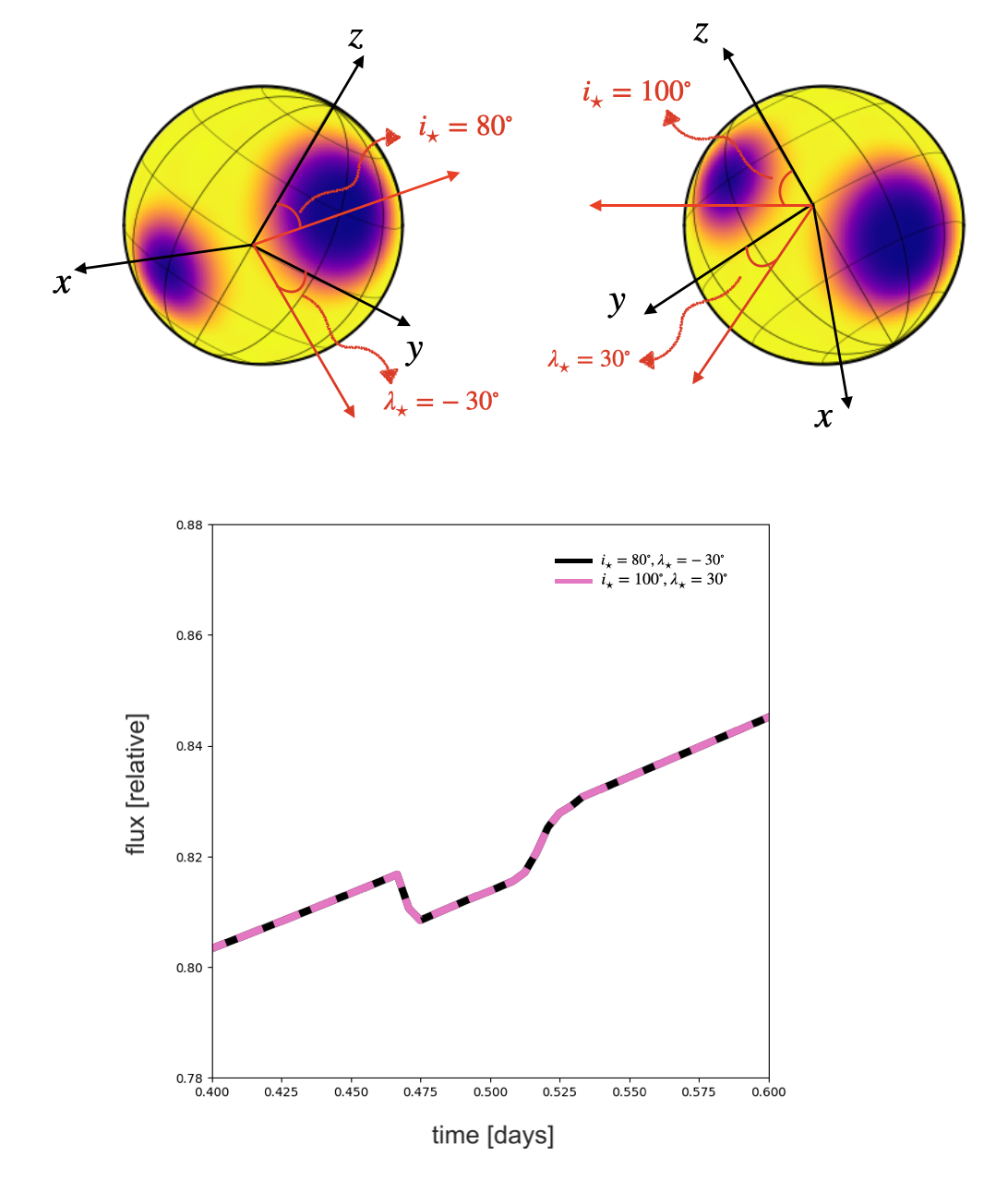}
        \caption{
            Illustration of the degeneracy between stellar inclination ($i_\star$) and obliquity ($\psi_\star$) in transit light curves. 
            The left panel shows a star-planet system with inclination $i_\star$ and obliquity $-\psi_\star$, while the right panel 
            depicts a system with inclination $180-i_\star$ and obliquity $\psi_\star$ and with a flipped map. 
            Despite the different geometries, both configurations result in identical transit light curves, shown in the central plot. 
            The x-axis represents time relative to the transit center, and the y-axis shows normalized flux. 
            This degeneracy demonstrates that while transit light curves can constrain the relative alignment between stellar spin and planetary orbit, 
            they cannot distinguish between these mirrored configurations without additional information. 
            The angles are represented with $x$, $y$, and $z$ vectors.
        }
        \label{fig:inc-obliquity-degeneracy}
    \end{centering}
\end{figure}

For example, a star with an inclination of $80^\circ$ and an obliquity of $-30^\circ$ will present the same transit light curves 
as a star with an inclination of $100^\circ$ and an obliquity of $30^\circ$ with a flipped map (see Figure \ref{fig:inc-obliquity-degeneracy}). 
In both cases, the angle between the stellar spin axis and the planet's orbital axis is identical, just reflected about the plane of the sky.
This degeneracy highlights a limitation in our ability to uniquely determine stellar geometry solely from transit light curve analysis. 
While we can constrain the relative alignment between the stellar spin and planetary orbit with high precision, we cannot distinguish between these 
mirrored configurations without additional information or observations.

Given the degeneracy between stellar inclination and obliquity, a naive sampling approach could lead to unnecessary computational costs and potential 
convergence issues. To address this, we can leverage the fact that inclination and obliquity can be represented as vectors in three-dimensional 
space (see Figure \ref{fig:inc-obliquity-degeneracy}). This allows us to implement a more efficient sampling strategy that breaks the degeneracy 
while maintaining physical consistency.

The key insight is that the problematic degeneracy corresponds to a reflection symmetry in 3D space. Instead of sampling directly 
in the degenerate $(i_\star, \psi_\star)$ parameter space, we transform to an auxiliary coordinate system where this symmetry can be naturally 
broken by transforming to a coordinate system where the plane of reflection is orthogonal to one of the coordinate axes. We introduce transformed variables $\tilde{x}$, $\tilde{y}$, and $\tilde{z}$, on
which we place prior distributions. Specifically, we use normal priors for
$\tilde{x}$ and $\tilde{y}$, and a half-normal prior for $\tilde{z}$ to ensure
it remains positive. The half-normal prior on $\tilde{z}$ is the crucial element that breaks the reflection degeneracy—by restricting $\tilde{z}$ to 
be positive, we effectively choose one side of the reflection symmetry and eliminate the redundant parameter space. These priors are chosen to explore the
non-degenerate part of the parameter space effectively
as described below.


The transformation from these sampled variables to the physical space begins with:
\par
\begin{eqnarray}
    x & = & \tilde{x}, \\ 
    y & = & \frac{\sqrt{2}}{2} (\tilde{z} + \tilde{y}), \\
    z & = & \frac{\sqrt{2}}{2} (\tilde{z} - \tilde{y})
\end{eqnarray}

From these transformed cartesian coordinates, we can derive the
stellar inclination and obliquity:
\begin{eqnarray}
    i_\star & = & \arctan{\frac{y}{x}}, \\ 
    \psi_\star & = & \arccos{\frac{z}{\sqrt{x^2 + y^2 + z^2}}}
\end{eqnarray}
With these definitions, the degeneracy identified above
corresponds to $\tilde{z} \to - \tilde{z}$, so the half-normal prior restricts
the model to a single branch of the degeneracy.  The isotropic priors on
$\tilde{x}$, $\tilde{y}$, and $\tilde{z}$ induce appropriate isotropic priors on
$i_\star$ and $\psi_\star$.

This transformation strategy addresses the degeneracy problem in three important ways. First, it eliminates the redundant 
parameter space by construction -- the half-normal prior on $\tilde{z}$ ensures we only sample from one side of the symmetry. 
Second, it preserves the physical meaningfulness of our constraints by maintaining isotropic priors on the final parameters. 
Third, it improves computational efficiency by preventing the sampler from wasting time exploring degenerate regions where different 
parameter combinations yield identical model predictions.

By sampling in this transformed space and then mapping back to the physical
parameters, we ensure that our model explores the full range of physically
meaningful configurations without redundancy. This not only improves the
efficiency of our sampling but also provides unambiguous constraints on the
stellar geometry.

While our current framework focuses on extracting maximum information from photometric time series, breaking this degeneracy fully 
requires incorporating complementary observational constraints. Spectroscopic measurements, which when combined with 
the photometrically-derived rotation period, can independently constrain inclination angles. 
Asteroseismic observations offer even more powerful constraints, as oscillation mode frequencies and splittings 
can provide precise measurements of both stellar inclination and internal rotation profiles, 
effectively decoupling geometric and rotational effects. Multi-observable approaches would not only resolve 
the inclination-obliquity degeneracy but also enable more robust characterization of stellar magnetic field geometries 
and the three-dimensional structure of active regions. However, joint modeling of multi-technique data is outside the scope 
of this work. In Section \ref{sec:discussion}, we compare our derived $v \sin{i}$ to the spectroscopically-measured $v \sin{i}$.

\subsection{Model for evolving surfaces}
\label{sec:evolmodel}
Some active stars, particularly Sun-like stars, exhibit surface evolution on timescales as short as a single rotation period. For example, starspots can appear and dissipate within days or weeks, 
while chromospheric plages and coronal loops can flare up and decay on even shorter timescales. Such rapid evolution of surface structures violates 
the fundamental assumption of a static photosphere underlying the model described in Section \ref{sec:model}. 
Accounting for these time-dependent surface phenomena is crucial for accurately interpreting observations of active stars, as they can introduce significant 
time-varying distortions in the photometric light curves. 

Here, we describe an improved model that accounts for the surface evolution.

To account for the time-varying nature of the stellar surface maps, we have extended the model described in Section \ref{sec:model} to allow 
for a continuous evolution of the maps over 
time. We assume that the surface map transitions smoothly from one epoch to the next, with the spherical harmonic 
coefficients linearly interpolated between consecutive map epochs. 

Specifically, let $\pmb{y}_1$,
$\pmb{y}_2$, $\pmb{y}_3$, ... represent the spherical harmonic coefficient
vectors describing the surface maps at a sequence of epochs $\pmb{t}_1$,
$\pmb{t}_2$, $\pmb{t}_3$, ... For any time $\pmb{t}$ between two consecutive
epochs $\pmb{t}_i$ and $\pmb{t}_{i+1}$, the coefficient vector $\pmb{y(t)}$ is
obtained by linear interpolation:
\begin{linenomath}\begin{align}\label{linearinterp}
    \mathbb{y}(t) = (1 - \alpha)\mathbb{y}_i + \alpha\mathbb{y}_{i+1},
\end{align}\end{linenomath}

where $\alpha = (t - t_i) / (t_{i+1} - t_i)$ is the linear interpolation factor between the epochs. 
This interpolation scheme ensures the surface map smoothly deforms from $\mathbb{y}_i$ at time $t_i$ to $\mathbb{y}_{i+1}$ at the next epoch $t_{i+1}$.

A key aspect is that each pair of consecutive maps $\mathbb{y}_i$, $\mathbb{y}_{i+1}$
is interpolated independently from other map pairs. Thus, $\mathbb{y}_{i+1}$ does
not depend on $\mathbb{y}_{i-1}$, and $\mathbb{y}_{i+2}$ is independent of
$\mathbb{y}_{i+1}$, allowing the model greater flexibility to fit arbitrary time
variations. The framework developed for static maps in Section \ref{sec:model}
is applied independently to each interpolated map $\mathbb{y}(t)$, with the
transition between maps governed by the linear interpolation above.
The predicted true flux remains linear in the maps; the complete
design matrix in this model is blocked and composed of values from the
constant-map design matrices produced by \texttt{starry} weighted by the linear
interpolation factors at the corresponding times.

There are two other assumptions to note here. First, the prior we place on each
of the maps $\pmb{y(t)}$ is a fresh draw from the \texttt{StarryProcess}.
Second, the linear interpolation approach is sensible if the $\Delta t$ between
the fitted maps is small enough. In Section \ref{sec:experiment2}, we assume the
spot time evolution is $\sim P_{\star}$. In general, we recommend to have the
epochs of the interpolated light
curve span times that are comparable to the spot evolution
timescale.

\section{Results: synthetic datasets}
\label{sec:results-synthetic}
\subsection{Priors}
Before presenting the results, it is essential to elucidate the prior
distributions employed throughout this work. We have implemented multiple priors
in this work, carefully selected to maintain a balance between informativeness
and generalizability. These priors serve to encapsulate our existing knowledge
and assumptions about the parameters of interest, while allowing for sufficient
flexibility to avoid computational cost.  We write
\begin{itemize}
    \item $\sim\mathcal{U}(\rm low, \rm high)$: a uniform distribution.
    \item $\sim \rm Log \mathcal{U}(\rm low, \rm high)$: a log-uniform distribution.
    \item \textit{Planetary Inclination}: Considering that $i_p=\arccos{b}$,
    where $b$ is the impact parameter, the prior on the inclination is defined
    via the uniform prior $\sim\mathcal{U}(-b_{max}, -b_{max})$ on $b$, where
    $b_{max} = R_{\star} / a$. This choice reflects the geometric constraint that physically meaningful transits 
    require $|b| \leq 1$ and accounts for the typical range of impact parameters observed in transiting exoplanet systems.
    \item \textit{Period}: a prior on $P$ and $P_{\star}$. Instead of directly
    constraining the period, this prior works in log-space. The prior is a uniform distribution in
    log-space, bounded by the log of the true value plus or minus some fraction. Log-uniform priors are particularly 
    appropriate for periods because they span multiple orders of magnitude and avoid biasing toward longer periods that 
    would result from uniform priors in linear space. This choice is well-motivated for both planetary orbital periods (which can range 
    from hours to years) and stellar rotation periods (which can range from days to months).
    \item \textit{Stellar Angle}: a normal distribution on orientation vectors $\tilde{x}$ and $\tilde{y}$ and a half-normal distribution 
    on $\tilde{z}$. See Section
    \ref{sec:obl-inc} for the full explanation. These priors are designed to produce isotropic distributions on the stellar 
    inclination and obliquity angles while breaking the inherent degeneracy between these parameters and implementing the correct spherical topology so the sampler experiences similar physical configurations as nearby points in the parameter space, as discussed in detail in Section \ref{sec:obl-inc}.
\end{itemize}

\subsection{Experiment I: non-evolving surface}
\label{sec:experiment1}
\begin{figure*}[ht!]
    \begin{centering}
        \includegraphics[width=\linewidth]{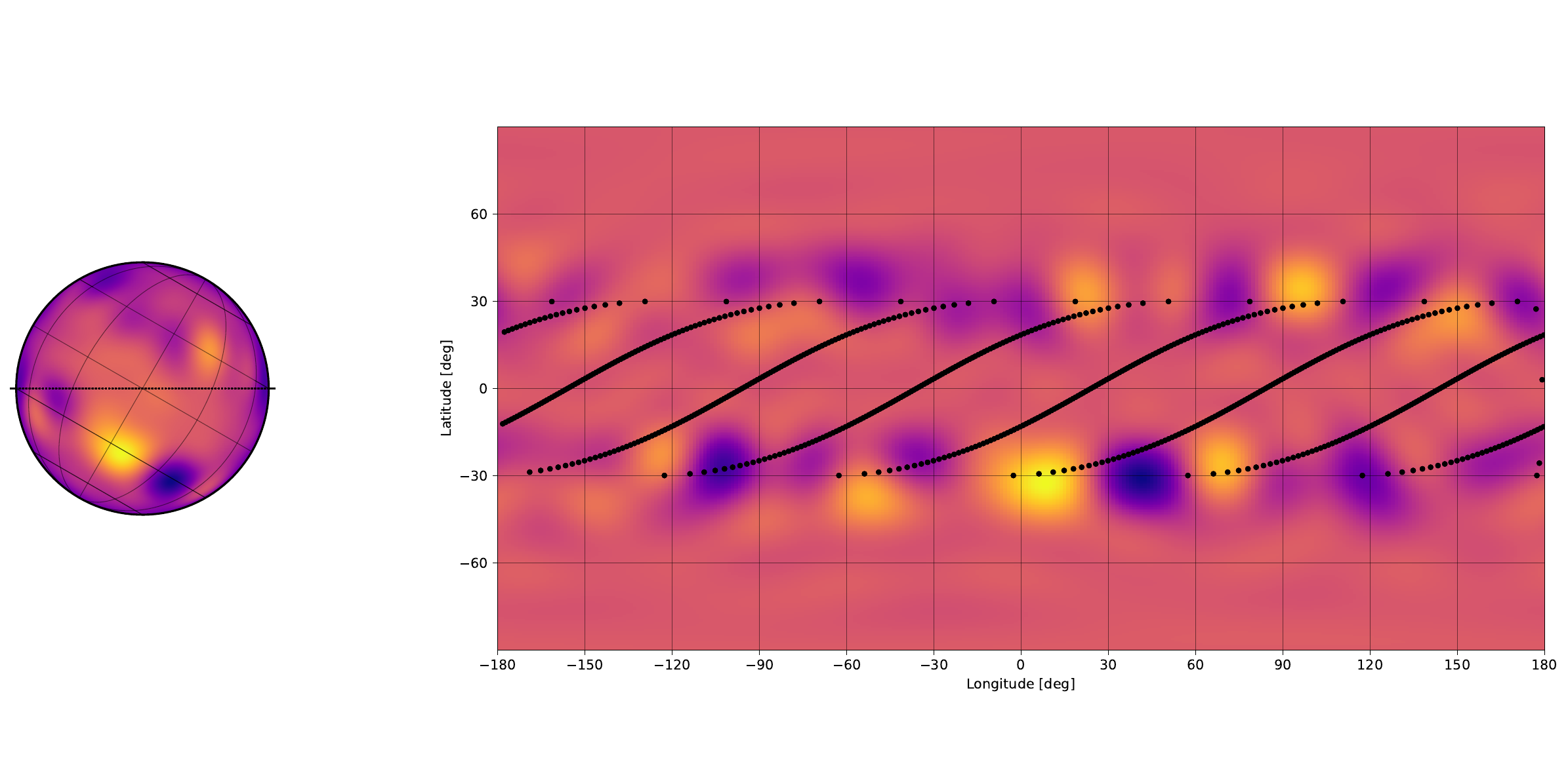}
        \includegraphics[width=\linewidth]{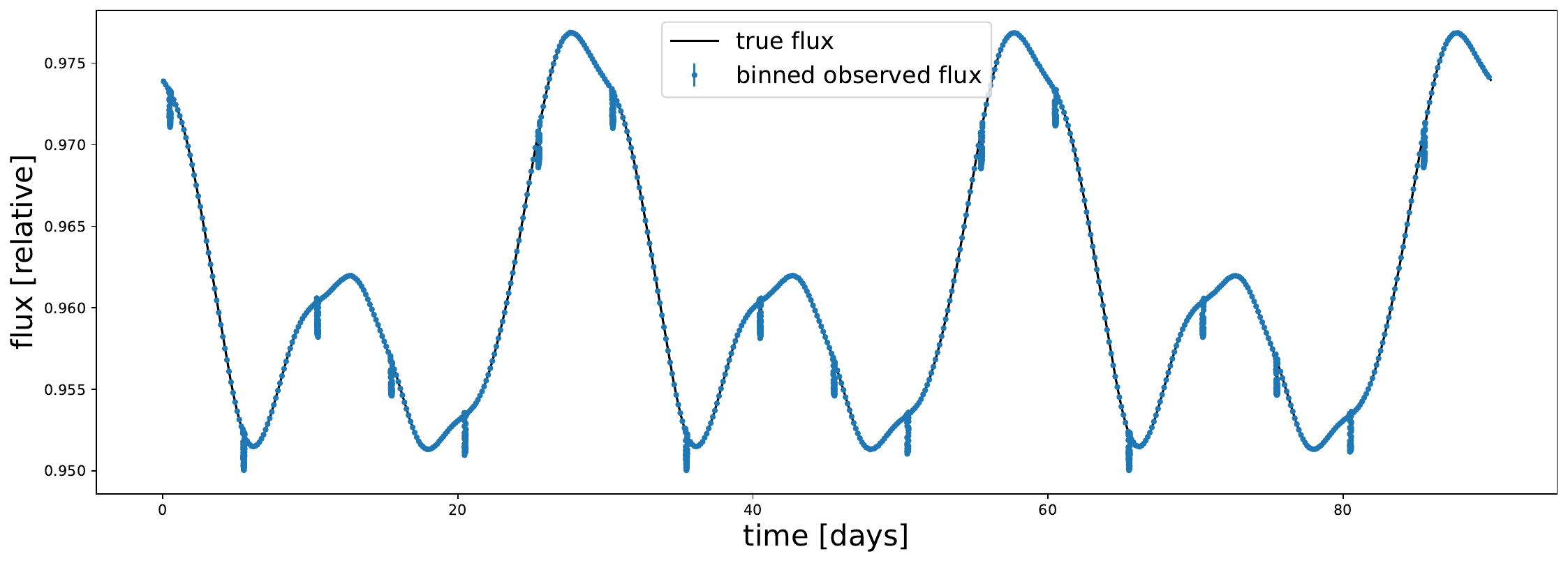}
        \caption{
            The upper panel shows the \textit{true} map for the Experiment I of this paper. The black dots on the map indicate the trajectory of the 
            planet as it moves across the face of the star during the transit event. The left panel 
            shows the stellar map along with its orientation on the sky and the right panel shows the rectangular projection of the map. 
            The bottom panel is the generated light curve. The black line is the \textit{true} light curve and the blue dots show the binned light curve
            that's used in the model.
            \href{https://github.com/ssagynbayeva/polka-dotted-stars-toi3884/blob/main/src/tex/notebooks/experiment-1.ipynb}{\faGithub}
            \href{https://zenodo.org/records/16647404}{\faDatabase}}
        \label{fig:experiment-1-map}
    \end{centering}
\end{figure*}

To verify the proper calibration of our model, we evaluate its performance using synthetic data, which is created through the following process.
We create a system consisting of a star and a planet, each of them have some orbital parameters presented in Table \ref{tab:LongPriors}. 
The \textit{true} map is just a random draw from the process evaluated up to spherical harmonic degree
$l_{max} = 15$ and conditioned on different values of the hyperparameter vector $\pmb{\theta}_\bullet$ (Table \ref{tab:LongPriors}). Figure \ref{fig:experiment-1-map} 
illustrates the setup of Experiment I. The upper panel displays the "true" map used in the experiment, which represents the underlying reality that 
the model aims to reconstruct. This map represents the brightness distribution across the star's surface. We generated 90 days of observations with a 2-minute
cadence, and then we binned the out-of-transit flux to speed up the sampling process.
Maps computed at lower spherical harmonic order, e.g., $l_{max} = 15$, are less computationally expensive to compute, but small 
$l_{max}$ cannot produce small scale surface brightness variations. For example, here, $l_{max} = 15$ will
only be able to produce the spots of sizes of $\sim 10^\circ$, which are bigger than any sunspots. 
We'll discuss the compromise between surface 
resolution and computational expense in the Section \ref{sec:smallspots}. 

We used No-U-Turn Sampling, a variant of Hamiltonian Monte Carlo \citep[NUTS;][]{Duane1987,Hoffman2011} to do posterior inference on our synthetic dataset.

We use several parameters in our sampling process: $i_p$ (the planetary inclination), $e$ (its eccentricity), $P$ (its orbital period),
$t_0$ (start of the transit time), $R_p/R_\star$ (ratio of planetary to stellar radius), $u_1$ and $u_2$ (limb darkening coefficients),
$i_\star$ (stellar inclination), $\psi_\star$ (stellar obliquity), $P_\star$ (stellar rotational period), $\mathbb{n}$ (the number of spots), 
$\mathbb{c}$ (their contrast), $\mathbb{r}$ (their radius), and $\mathbb{a}$ and $\mathbb{b}$ (parameters of the Beta distribution that describe 
how the spots are spread across latitudes).  

\begin{figure*}[hbt!]
    \begin{centering}
        \includegraphics[width=\linewidth]{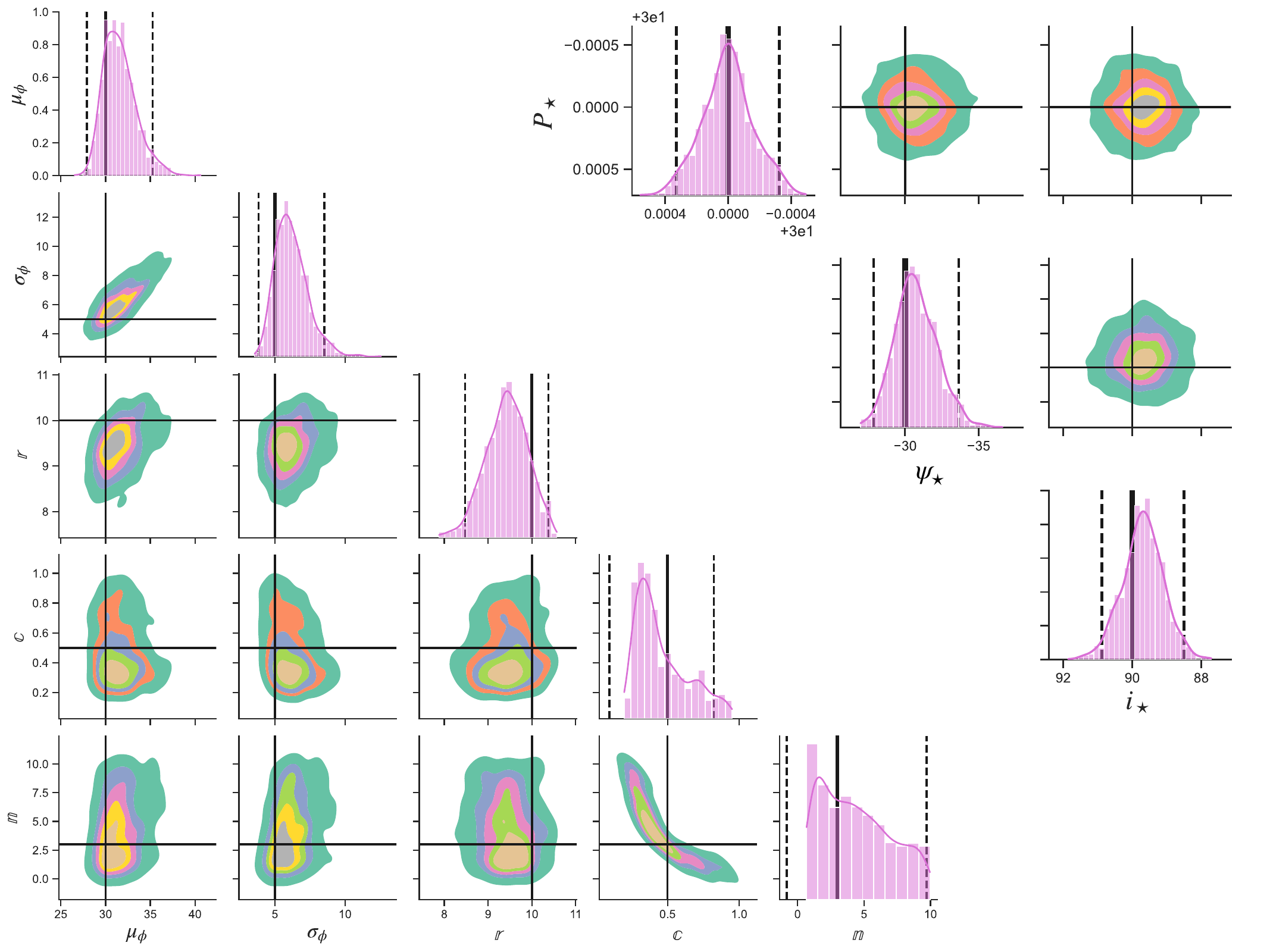}
        \caption{
            Posterior distributions for the all the parameters $\pmb{\Theta}$ for the synthetic data run 
            (Table \ref{tab:LongPriors} and Figure \ref{fig:experiment-1-map}). The axes span the entire prior volume, and 
            the black lines indicate the true (input) values, the dashed black lines indicate 2 standard deviations.
            In our analysis process, 
    when we encounter a solution that gives the obliquity $\psi_{\star, \rm inferred}$ instead of $-\psi_{\star, \rm inferred}$, we systematically transform the results. 
    Specifically, we change the inclination to $180^\circ-i_{\star, \rm inferred}$ and obliquity to $-\psi_{\star, \rm inferred}$ for visualization purposes. 
    \href{https://github.com/ssagynbayeva/polka-dotted-stars-toi3884/blob/main/src/tex/notebooks/experiment-1.ipynb}{\faGithub}
    \href{https://zenodo.org/records/16647404}{\faDatabase}}
        \label{fig:experiment-1-map-corner-all}
    \end{centering}
\end{figure*}

We use Equation \ref{eq:log-likeRodrigo} as our log likelihood term. The results
are shown on Figure \ref{fig:experiment-1-map-corner-all}, where we correctly
infer almost all five starspot parameters within $\sim 2$ standard deviations. 

\begin{figure*}[hbt!]
    \begin{centering}
        \includegraphics[width=\linewidth]{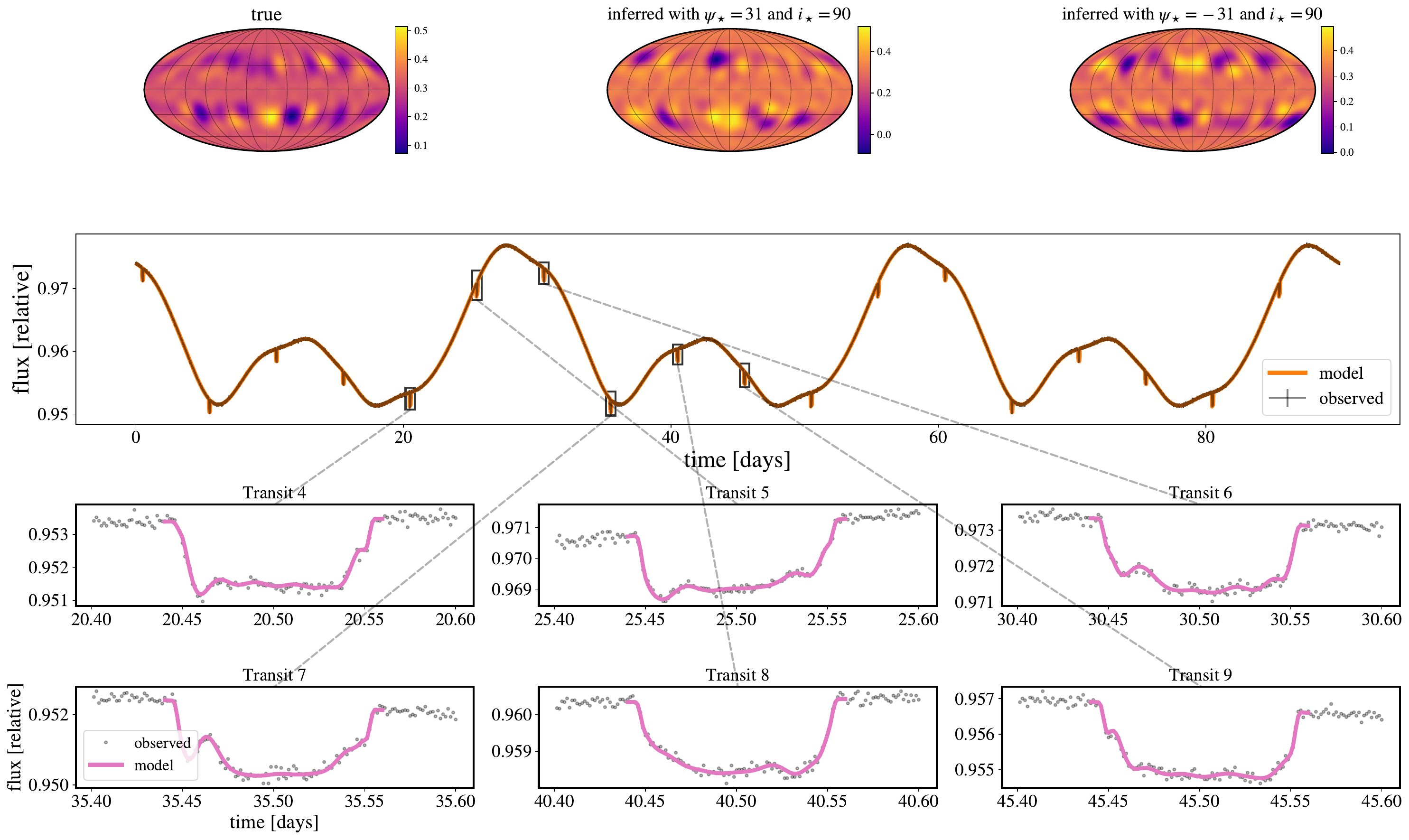}
        \caption{
            Comparison of true and inferred stellar surface maps, light curve fit, and transit details. Top panel: the true stellar surface map used to 
            generate synthetic data and two inferred stellar surface maps due to the obliquity-inclination degeneracy, representing the mean of 1000 posterior samples. Color scales represent spot 
            intensity. The similarity between true and inferred maps demonstrates the model's capability to recover the overall spot distribution. Bottom panel:
            Full light curve comparison. Black points show the synthetic data with added noise. The orange line represents the best-fit model, 
            computed as the mean of 1000 posterior samples. Bottom panels (three rows): zoom-in views of six consecutive transits. Black points are data, 
            pink lines show the best-fit model. Note the spot-crossing events visible as brief brightness increases during transits. 
            The light curve fit shows strong agreement with the data over the full time series, while the transit close-ups highlight the model's ability 
            to capture subtle spot-crossing events and their evolution over consecutive transits.
            \href{https://github.com/ssagynbayeva/polka-dotted-stars-toi3884/blob/main/src/tex/notebooks/experiment-1.ipynb}{\faGithub}
            \href{https://zenodo.org/records/16647404}{\faDatabase}}
        \label{fig:experiment-1-map-comparison}
    \end{centering}
\end{figure*}

\begin{figure*}[hbt!]
    \begin{centering}
        \includegraphics[width=\linewidth]{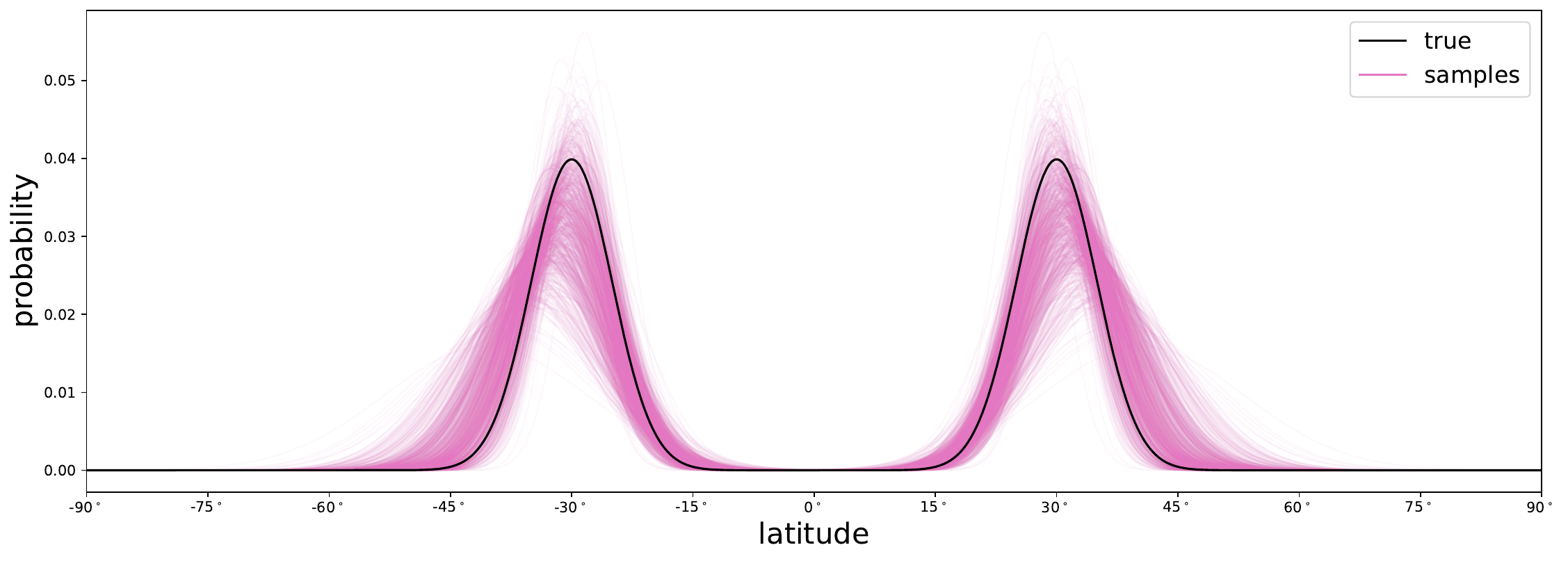}
        \caption{
            Posterior distributions of spot latitudes derived from synthetic data analysis. The pink curves represent individual Beta distribution 
            probability density functions (PDFs) for spot latitude, each generated using parameters sampled from the posterior distributions of 
            $\mu_\phi$ and $\sigma_\phi$ (Figure \ref{fig:experiment-1-map-corner-all}). The ensemble of pink curves illustrates the range of 
            plausible spot latitude distributions, effectively quantifying our inferred knowledge about spot distribution patterns across the stellar surface 
            in our dataset. The black curve shows the true underlying distribution used to generate the synthetic spot data (parameters given in Table 
            \ref{tab:LongPriors}). Note the general agreement between the inferred distributions and the true distribution, 
            demonstrating the model's ability to recover the input spot latitude distribution within the uncertainties of the inference process.
            \href{https://github.com/ssagynbayeva/polka-dotted-stars-toi3884/blob/main/src/tex/notebooks/experiment-1.ipynb}{\faGithub}
            \href{https://zenodo.org/records/16647404}{\faDatabase}
        }
        \label{fig:experiment-1-acttive-lats}
    \end{centering}
\end{figure*}

To visually demonstrate the performance of our model, Figure \ref{fig:experiment-1-map-comparison} presents a comprehensive comparison between 
the true stellar surface used in our synthetic data generation and the inferred surface map from our analysis, alongside the resulting light curve fit and 
transit details. Of particular interest are the three consecutive transits shown in detail on the right panel. These zoomed-in views highlight our 
model's ability to precisely fit transit events, including subtle spot-crossing phenomena. The spot-crossing events, visible as brief increases in brightness 
during the transits, are well-captured by our model. It can be seen that true and inferred maps are not \emph{exactly} the same, but share the same surface features -- i.e. bright spots at the
same locations, same active latitudes, etc. -- which still explains the surface of the star.
To evaluate the effectiveness of our model in recovering the underlying spot distribution, we examined the posterior distributions of spot latitudes 
derived from our synthetic data analysis. Figure \ref{fig:experiment-1-acttive-lats} illustrates these results, and the constraints we can place on 
stellar spot distributions. In Figure \ref{fig:experiment-1-acttive-lats}, each pink curve represents a Beta distribution PDF for spot latitude. 
These curves are generated using parameters sampled from the posterior distributions of $\mu_\phi$ and $\sigma_\phi$, which characterize the 
latitude distribution in our model. The collection of pink curves effectively creates a distribution that quantifies our inferred knowledge about 
how spots are distributed across the stellar surface in our dataset. The variability among these pink curves reflects the uncertainty in our inference. 
Regions where the curves are more tightly clustered indicate greater certainty in our estimates, while areas with more spread suggest higher uncertainty. 
This visualization allows us to assess not just a single "best-fit" distribution, but the full range of plausible distributions consistent with our data and model.
The black curve in Figure \ref{fig:experiment-1-acttive-lats} represents the true underlying distribution used to generate the synthetic spot data, 
with parameters as specified in Table \ref{tab:LongPriors}. The general agreement between the ensemble of pink curves and this 
true distribution is encouraging, demonstrating that our model can successfully recover the input spot latitude distribution within the 
uncertainties of the inference process.

The parameters listed in Table \ref{tab:LongPriors} can be grouped into three categories. 
\textit{Transit parameters} describe the planetary system geometry and include the planetary inclination ($i_p$), 
eccentricity ($e$), orbital period ($P$), time of transit center ($t_0$), planet-to-star radius ratio ($R_p/R_\star$), and 
limb darkening coefficients ($u_1$, $u_2$). \textit{Stellar orientation parameters} characterize the star's rotational state and 
include the stellar inclination ($i_\star$), obliquity ($\psi_\star$), and stellar rotation period ($P_\star$). \textit{Starspot parameters} 
describe the surface features and include the characteristic spot size ($\mathbb{r}$), spot contrast ($\mathbb{c}$), the number of spots 
($\mathbb{n}$), and the parameters controlling the latitudinal distribution of spots ($\mu_\phi$ and $\sigma_\phi$, which represent 
the mean and variance of the spot latitude distribution, respectively).

\begin{table}[]
    \vspace{0.5cm}
    \centering
    \caption{The free parameters for the synthetic data in Section \ref{sec:experiment1}, their true values, and their priors.}
    \begin{tabular}{llll}
    \hline
    Parameter                                 & True value            & Prior distribution                   & Inferred value    \\ \hline\hline
    $i_p$                                     & $90^\circ$            & Planetary Inclination                & $90.04^{+0.104}_{-0.089}$                   \\
    $e$                                       & $0.2$                 & $\sim\mathcal{U}(0, 0.5)$            & $0.224 \pm 0.001$                     \\
    $P$                                       & $5$ days              & Period                               & $5.0^{+7\times10^{-6}}_{-6.45\times10^{-6}}$                       \\
    $t_0$                                     & $0.5$                 & $\sim\mathcal{U}(-0.6, 0.6)$         & $0.5^{+10^{-4}}_{-9.8\times10^{-5}}$                       \\
    $R_p / R_\star$                           & $0.04$                & $\sim \rm Log \mathcal{U}(0.02, 0.08)$ & $0.04 \pm 2\times10^{-4}$                         \\ 
    $u_1$                                     & $0.4$                 & $\sim\mathcal{U}(0, 0.6)$            & $0.38^{+0.07}_{-0.06}$                         \\ 
    $u_2$                                     & $0.26$                & $\sim\mathcal{U}(0, 0.4)$            & $0.29 \pm 0.06$                         \\ \hline
    $i_\star$                                 & $90^\circ$            & Stellar Angle                        & $89.68^{+0.6}_{-0.56}$                      \\
    $\psi_\star$                           & $-30^\circ$           & Stellar Angle                        & $-30.67^{+1.33}_{-1.48}$                           \\
    $P_\star$                                 & $30$ days             & Period                               & $30\pm 2 \times10^{-4}$                       \\ \hline
    $\mathbb{r}$                              & $10^\circ$            & $\sim\mathcal{U}(5, 25)$             & $9.44^{+0.447}_{-0.497}$                           \\
    $\mathbb{c}$                              & $0.5$                 & $\sim\mathcal{U}(0.01, 1)$           & $0.46^{+0.28}_{-0.1}$                               \\
    $\mathbb{n}$                              & $3$                   & $\sim\mathcal{U}(0, 10)$             & $4.45^{+3.53}_{-2.6}$                               \\
    $\mu_\phi$                                & $30^\circ$            & $\sim\mathcal{U}(0, 1)$ on $\mathbb{a}$ and $\mathbb{b}$ & $31.6^{+1.99}_{-1.52}$                               \\
    $\sigma^2_\phi$                           & $5^\circ$             & $\sim\mathcal{U}(0, 1)$ on $\mathbb{a}$ and $\mathbb{b}$ & $6^{+1.2}_{-0.95}$                      \\ \hline
    \label{tab:LongPriors}
    \end{tabular}
\end{table}

It is important to note that all results presented in this analysis have been carefully adjusted to account for the inclination and obliquity 
degeneracy discussed in Section \ref{sec:obl-inc}. As discussed earlier, a system with stellar inclination $i_\star$ and obliquity $-\psi_\star$ 
produces identical light curves to a system with inclination $180^\circ-i_\star$ and obliquity $\psi_\star$. In our analysis process, 
when we encounter a solution that gives the obliquity $\psi_\star$ instead of $-\psi_\star$, we systematically transform the results. 
Specifically, we change the inclination to $180^\circ-i_\star$ for visualization purposes in Figure \ref{fig:experiment-1-map-corner-all}. 
This approach ensures consistency in our reported results and allows for meaningful comparisons across different systems or models. 
By explicitly addressing this degeneracy, we provide a more robust and physically interpretable set of results, avoiding potential ambiguities in the 
orientation of the stellar spin axis relative to the planetary orbit. To show this explicitly, in Figure \ref{fig:experiment-1-map-comparison}
we plot the inferred maps, where the middle one is the map with the degeneracy accounted for, and the right one is the map with the actual inferred values 
for obliquity and inclination (degeneracy is not accounted for). The two maps are inverses of each other.

Finally, our tests demonstrated that attempting to model fine-scale features ($l_{max}=15$ spots) using lower spherical 
harmonic orders ($l_{max}=6$) produced inaccurate reconstructions. However, active latitudes would still be recoverable, 
though the reconstructed spots would appear larger than in reality. This size distortion consequently affects both the contrast 
and the apparent number of spots in the model. Despite these constraints, the approach aligns well with observations of 
many young and active stars, where extensive spot coverage is common. 

While our synthetic data validation demonstrates the model's performance under somewhat idealized conditions, real astronomical 
observations present additional challenges including irregular sampling, data gaps, instrumental systematics, and varying noise levels. 
Our model is inherently robust to many of these observational constraints due to its underlying mathematical framework and implementation.
Data gaps and irregular sampling, which are common in space-based photometry missions like TESS, do not fundamentally compromise the model's 
performance. This robustness is demonstrated in our analysis of the TESS example in Section \ref{sec:toi3884}, where we successfully recovered 
stellar parameters despite the presence of gaps in the observational coverage. The Gaussian process framework naturally handles 
irregularly spaced data points without requiring interpolation or gap-filling procedures. When data gaps occur -- even if they result in 
missing entire transit events -- the model simply adjusts its parameter inference based on the available observations. The probabilistic nature 
of our approach means that missing data points are effectively ignored rather than treated as zero-flux measurements, allowing the model 
to maintain its predictive accuracy across the observed time series. If a gap in the data coincides with a predicted transit time, 
the model does not penalize this absence but rather uses the remaining transit observations to constrain the system parameters.

\subsection{Experiment II: evolving surface}
\label{sec:experiment2}

While our primary analysis assumes a static spot distribution across the observational baseline, stellar magnetic activity is 
inherently dynamic. Active regions on stars like TOI-3884 evolve over time, with spots emerging, changing, and disappearing on 
timescales of weeks to months. To explore this time dimension, we developed a preliminary model incorporating temporal 
evolution of the stellar surface features. This experimental approach represents a conceptual extension of our methodology 
rather than a fully validated model, yet it offers compelling insights into the potential for tracking stellar surface evolution 
through transit observations.

By allowing key spot parameters to vary as functions of time between transit epochs, we can begin to probe questions about spot 
lifetimes, differential rotation, and activity cycles—aspects of stellar physics typically requiring long-term monitoring campaigns. 
Our evolving surface model maintains the core framework of our primary analysis while introducing time-dependence to the 
spot distribution through a Gaussian process temporal kernel. This creates a coherent evolution of the surface features that 
respects both the constraints from individual transit events and the physical expectations for how stellar spots behave.
Though this approach requires additional assumptions and computational complexity, initial results suggest that transit light 
curves do contain retrievable information about surface feature evolution. The methodology presented here offers a 
promising avenue for future work, potentially enabling new types of stellar activity studies using already-available 
transit photometry from missions like TESS, Kepler, and the upcoming PLATO mission.

\begin{figure*}[hbt!]
    \begin{centering}
        \includegraphics[width=\linewidth]{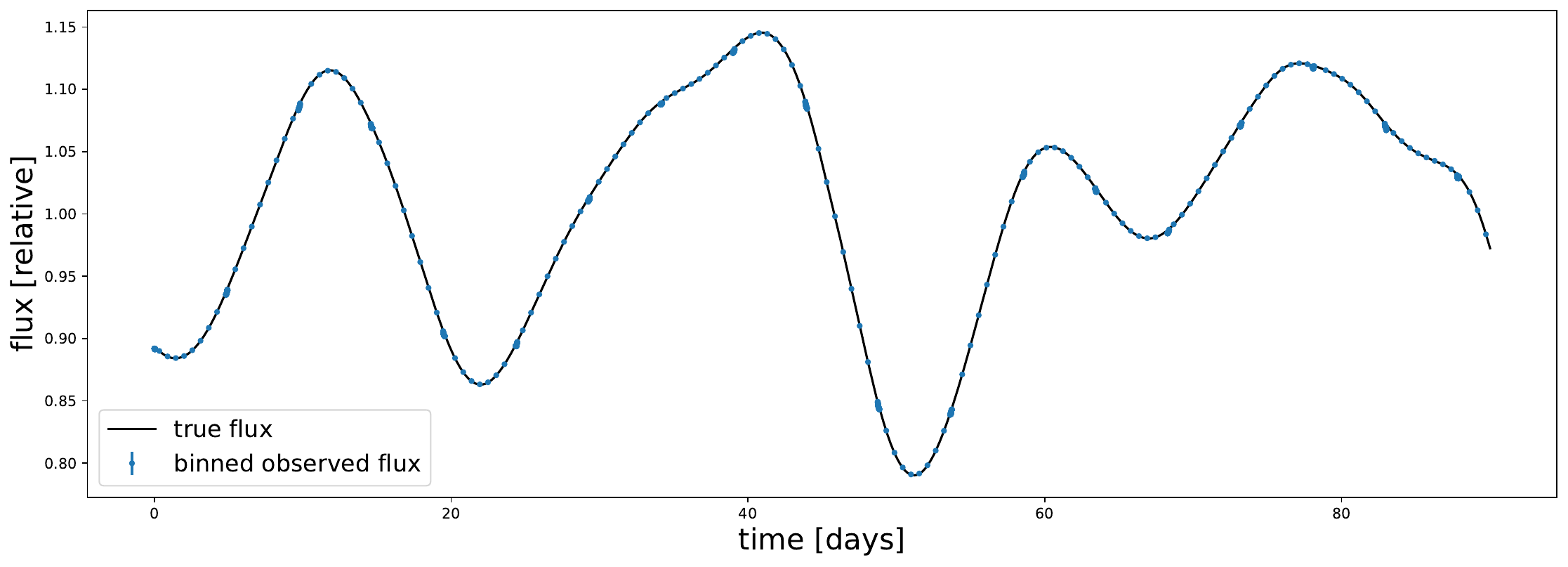}
        \caption{
            Simulated light curve from our evolving stellar surface model spanning approximately 150 days. The black line represents 
            the true underlying flux variations, while the blue points with error bars show binned simulated observations. 
            The complex pattern of photometric variability, with amplitude variations between approximately 0.87 and 1.12 in 
            relative flux, reflects the combined effects of stellar rotation and evolving surface features. Note the non-periodic 
            nature of the variations and changing amplitude over time, indicative of multiple active regions evolving at different 
            rates and locations on the stellar surface.
            \href{https://github.com/ssagynbayeva/polka-dotted-stars-toi3884/blob/main/src/tex/notebooks/experiment-2.ipynb}{\faGithub}
            \href{https://zenodo.org/records/16647404}{\faDatabase}
        }
        \label{fig:experiment-2-lc}
    \end{centering}
\end{figure*}

\begin{figure}[hbt!]
    \begin{centering}
        \includegraphics[width=0.86\linewidth]{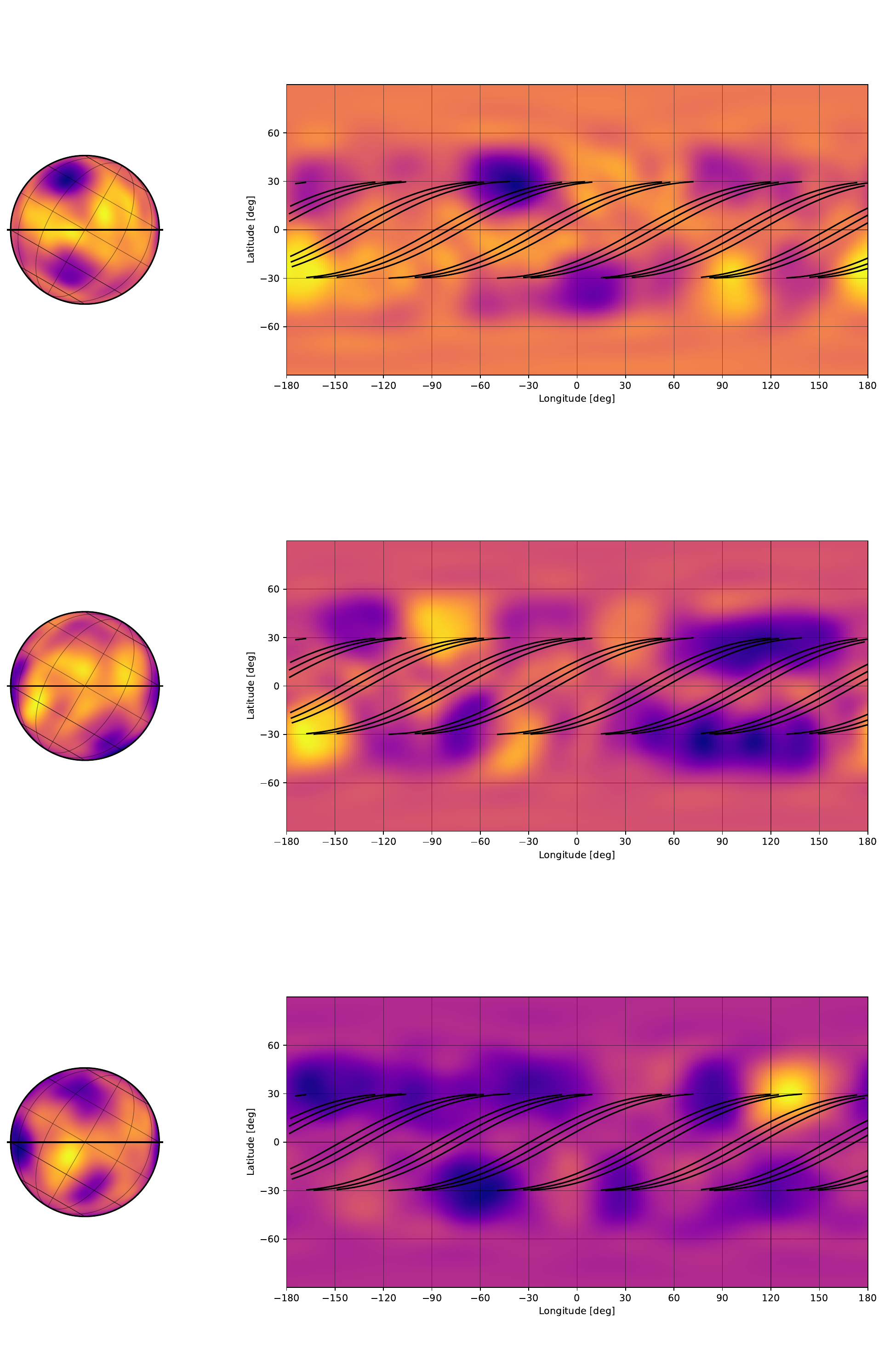}
        \caption{Simulated stellar surface maps at three representative epochs corresponding to the light curve in 
        Figure \ref{fig:experiment-2-lc}. Each row presents both a 3D visualization of the visible stellar hemisphere 
        (left) and a Mollweide projection of the complete surface (right). Dark purple regions indicate cooler spots, while 
        yellow-orange areas represent the warmer photosphere. The maps illustrate the evolution of active regions over time, 
        including changes in spot distribution, emergence of new features, and potential differential rotation effects. 
        The dotted lines on the 3D visualizations indicate the planet's transit path. These surface maps demonstrate how the 
        complex photometric variations in the light curve arise from the dynamic nature of stellar activity patterns.
        \href{https://github.com/ssagynbayeva/polka-dotted-stars-toi3884/blob/main/src/tex/notebooks/experiment-2.ipynb}{\faGithub}
        \href{https://zenodo.org/records/16647404}{\faDatabase}
    }
        \label{fig:experiment-2-true-maps}
    \end{centering}
\end{figure}

Figures \ref{fig:experiment-2-lc} and \ref{fig:experiment-2-true-maps} showcase the experimental results from our 
evolving surface model. Figure \ref{fig:experiment-2-lc} presents a simulated light curve spanning approximately 150 days, 
demonstrating how stellar rotation and evolving surface features produce complex photometric variations. The black line 
represents the true underlying flux variations, while the blue points with error bars show binned simulated observations. 
This light curve exhibits amplitude variations between approximately 0.87 and 1.12 in relative flux, with complex patterns 
that cannot be explained by simple periodic modulation, indicating the presence of multiple evolving active regions.

Figure \ref{fig:experiment-2-true-maps} complements this time-series data by displaying the corresponding stellar 
surface maps at three representative epochs. Each row presents both a 3D visualization of the stellar hemisphere facing the 
observer (left panels) and a Mollweide projection of the complete stellar surface (right panels). 
These maps reveal the dynamic nature of the spot distribution, with active regions emerging, evolving, and disappearing over time
from top to bottom. Particularly notable is the migration of spots in both latitude and longitude, suggesting potential 
differential rotation, as well as changes in the size and intensity of individual features.


\begin{figure*}[hbt!]
    \begin{centering}
        \includegraphics[width=\linewidth]{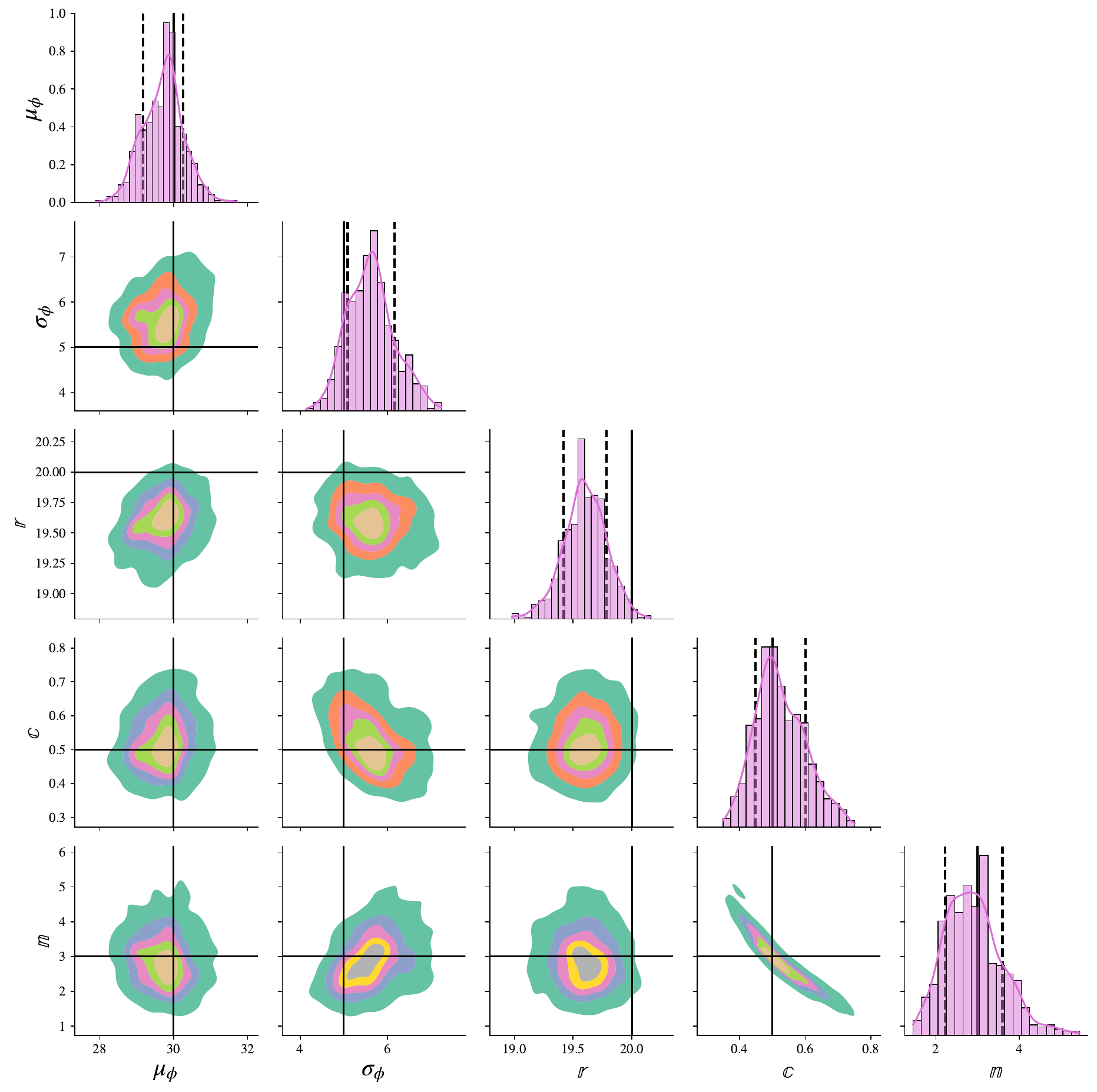}
        \caption{Posterior distributions and correlations for key spot model hyperparameters recovered from simulated data. 
        The solid black lines indicate true parameter values used to generate the simulated light curves, while dashed black 
        lines show the $\pm1\sigma$ boundaries. Parameters include spot latitude mean ($\mu_\phi$), latitude standard deviation 
        ($\sigma_\phi$), radius ($\mathbb{r}$), spot contrast ($\mathbb{c}$), and number of spots ($\mathbb{n}$). 
        The model successfully recovers all parameters within the expected uncertainty ranges.
         Note the correlation between number of spots and contrast, a well-known degeneracy in spot modeling.
         \href{https://github.com/ssagynbayeva/polka-dotted-stars-toi3884/blob/main/src/tex/notebooks/experiment-2.ipynb}{\faGithub}
         \href{https://zenodo.org/records/16647404}{\faDatabase}
    }
        \label{fig:experiment-2-corner}
    \end{centering}
\end{figure*}

\begin{figure*}[hbt!]
    \begin{centering}
        \includegraphics[width=\linewidth]{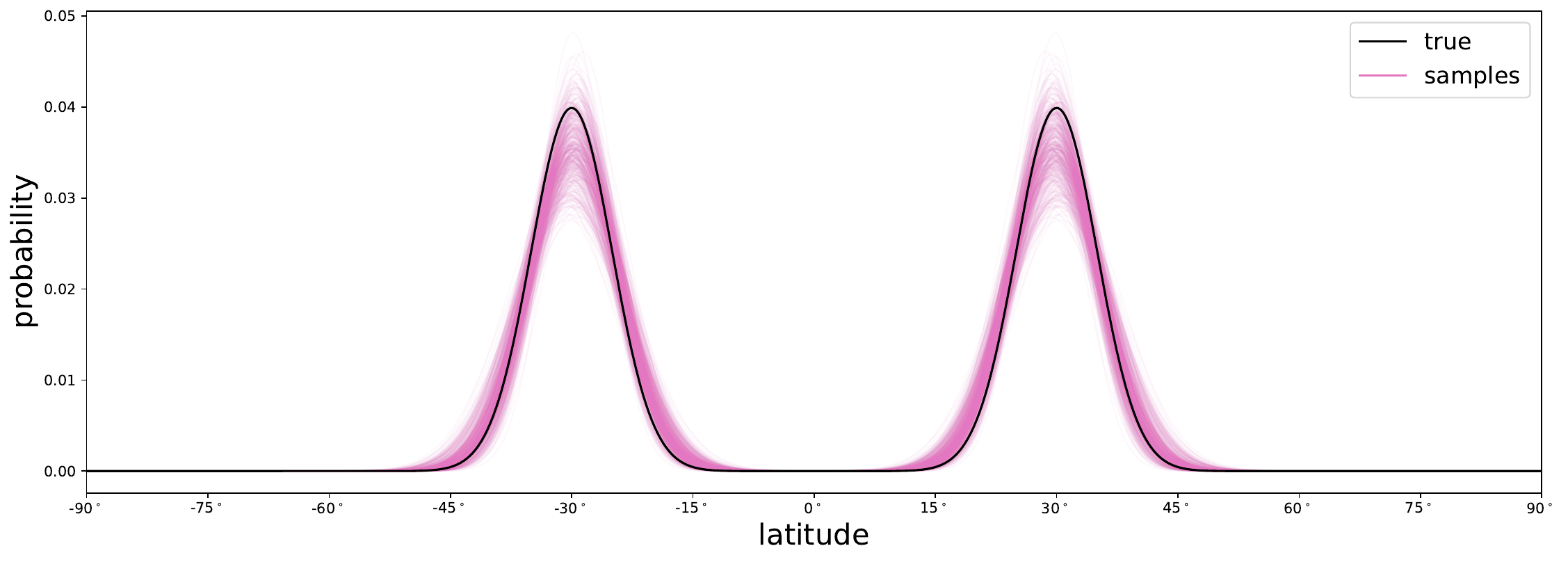}
        \caption{Posterior probability distribution of spot latitudes for Section \ref{sec:experiment2} just like the Figure 
        \ref{fig:experiment-1-acttive-lats}. The black line shows the true distribution, 
        while the pink lines represent individual posterior samples.
        \href{https://github.com/ssagynbayeva/polka-dotted-stars-toi3884/blob/main/src/tex/notebooks/experiment-2.ipynb}{\faGithub}
        \href{https://zenodo.org/records/16647404}{\faDatabase}
        }
        \label{fig:experiment-2-active}
    \end{centering}
\end{figure*}

Figure \ref{fig:experiment-2-corner} presents the posterior distributions for the key hyperparameters in our spot model. 
The parameters include the spot latitude mean ($\mu_\phi$), latitude variance ($\sigma^2_\phi$), spot contrast parameter 
($\mathbb{c}$), radius ($\mathbb{r}$), and the number of spots ($\mathbb{n}$).
The posterior for spot latitude mean ($\mu_\phi$) is tightly constrained around 30 degrees, consistent with 
the true value (shown with black lines). We see excellent recovery of the latitude mean, with the true value falling well 
within our posterior distribution and the 1$\sigma$ uncertainty bounds from the simulation aligning well with the width of 
our recovered posterior. The latitude variance ($\sigma^2_\phi$) is centered around 5-6 square degrees, again accurately 
capturing the true value indicated by the black line. Our model successfully recovers this aspect of the spot distribution, 
and our uncertainty estimates appear consistent with the true 1$\sigma$ range. The active latitudes are shown on Firgure \ref{fig:experiment-2-active}.
The radius ($\mathbb{r}$) parameter, which controls the typical size of spot features, is well-constrained at 
approximately 19.5 degrees. Future models will likely improve the recovery of this parameter by incorporating 
multi-wavelength transit observations, which would better constrain the physical extent of surface features by leveraging 
the wavelength-dependent contrast of spots.
The correlation between number of spots ($\mathbb{n}$) and contrast ($\mathbb{c}$) shows 
a well-known degeneracy between spot contrast and number of spots in spot modeling: as the number of spots increases, the contrast tends to decrease to 
maintain a similar overall flux effect.

\begin{figure*}[hbt!]
    \begin{centering}
        \includegraphics[width=\linewidth]{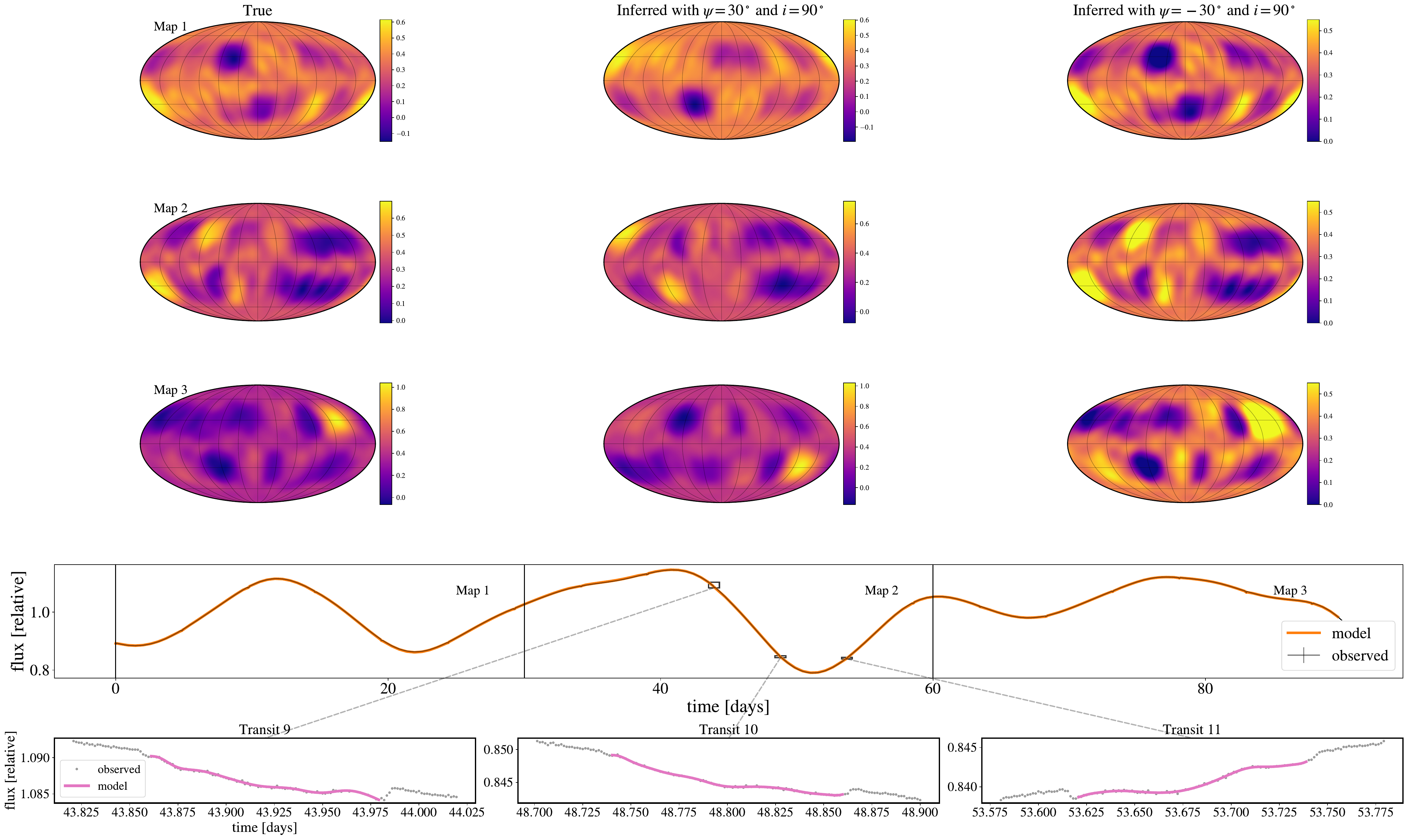}
        \caption{Comparison between true and inferred stellar surface maps from our evolving surface model using simulated data. 
        Top panel: True stellar surface maps (left column) 
        and our inferred reconstructions averaged over 1000 samples (middle column and the right column) at three epochs. Bottom: Full light curve 
        with vertical lines marking the three mapped epochs. The very last row shows three random consecutive transits. Note the north-south inversion in the inferred maps compared to 
        the true maps, demonstrating the inclination-obliquity degeneracy discussed in Section \ref{sec:obl-inc}. 
        Despite this expected degeneracy, our model successfully recovers the number, approximate size, and longitudinal 
        positioning of major active regions, with excellent agreement between the model light curve (red line) and simulated 
        observations (gray points).
        \href{https://github.com/ssagynbayeva/polka-dotted-stars-toi3884/blob/main/src/tex/notebooks/experiment-2.ipynb}{\faGithub}
        \href{https://zenodo.org/records/16647404}{\faDatabase}
        }
        \label{fig:experiment-2-results}
    \end{centering}
\end{figure*}

Figure \ref{fig:experiment-2-results} presents a comprehensive view of our evolving surface model's ability to reconstruct 
stellar surface features from simulated light curve data. The figure is organized into three key components that demonstrate 
the model's performance at different epochs.
The top portion shows the full simulated light curve (center panel), with three zoomed-in regions highlighting three random consecutive transits 
(top three panels). For each epoch, we compare the observed simulated data points (gray dots) with our model fit (pink line), 
demonstrating excellent agreement between the observations and our reconstruction.
The central portion of the figure displays the true stellar surface maps (left column) alongside our (averaged over 1000 samples) 
inferred reconstructions 
(middle column) at three different epochs, labeled as Map 1, Map 2, and Map 3, where Map 1 evolves to Map 2, which evolves to Map 3. 
Each map is shown as a Mollweide projection with a consistent color scale where dark purple regions indicate cooler spots and 
yellow-orange areas represent the warmer photosphere. The inferred maps show a north-south flip compared to the true maps. 
This hemisphere inversion is a direct manifestation of the inclination-obliquity degeneracy discussed in Section \ref{sec:obl-inc}, 
where transit observations alone cannot distinguish between certain combinations of stellar inclination and obliquity that result 
in mirrored surface feature distributions. Therefore, the resulting obliquity of the star is inferred to be 
$\psi_\star=30.49^{+0.58}_{-0.49}$ while the true obliquity is $-30^\circ$ (so, the opposite). 
Additionally, the stellar inclination's true 
values is $90^\circ$ while the inferred value is $i_\star=91.36^{+0.33}_{-0.36}$ (close to $180^\circ-i_{\rm true}$).
Despite this expected degeneracy, the inferred maps successfully capture the essential features of the true spot distribution, 
including the number, approximate size, and longitudinal positioning of major active regions. This is particularly evident in 
Map 3, where the model correctly identifies a prominent high-latitude bright region, though it places it in the opposite 
hemisphere.
The right column shows the corresponding light curve sections for each map epoch, further illustrating how well our model 
reproduces the observed flux variations despite the hemisphere ambiguity. The bottom panel presents the complete light curve 
with vertical lines marking the times corresponding to the three maps, contextualizing how the surface features evolve over the 
full observational baseline.
This figure demonstrates that while inherent degeneracies prevent unique determination of the absolute spot latitudes from 
transit data alone, our model can nevertheless recover detailed and physically meaningful information about the evolution of 
stellar surface features.

\section{The Case of TOI-3884}
\label{sec:toi3884}
In this section, we present the results of the model application on TESS light curves of TOI-3884. While TOI-3884 presents a relatively straightforward case, 
it serves as an excellent testbed for demonstrating the capabilities and functionality of our model. By applying our model to TOI-3884, we aim to validate 
its performance on a well-understood system, highlight its ability to extract meaningful information from a real light curve, 
and lay the groundwork for more complex analyses by establishing a baseline of performance.

TOI-3884 (also known as TIC 86263325) was observed by the TESS mission across multiple sectors. The star was monitored for 81 days over a 765-day span, 
using both 30-minute and 2-minute cadences. While Gaia observations show that the TESS aperture for this target isn't significantly contaminated by nearby stars, 
the light curve reveals several transiting spot crossings, indicating stellar activity. Various light curve products were generated for TOI-3884, 
including Simple Aperture Photometry (SAP) 
from the TESS Science Processing Operations Centre and Kepler Spline SAP (KSPSAP) from the Quick-Look Pipeline \citep{Huang2020}. 

TESS revisited TOI-3884 in Sectors 46 and 49, spanning parts of late 2021 and early 2022, using short-cadence observations with 2-minute exposures. 
We used the \texttt{lightkurve} package \citep{lightkurve} to process data from all three sectors, applying strict quality control measures to 
the Pre-search Data Conditioning Simple Aperture Photometry \citep{jenkins2016}.

\begin{figure*}[hbt!]
    \begin{centering}
        \includegraphics[width=\linewidth]{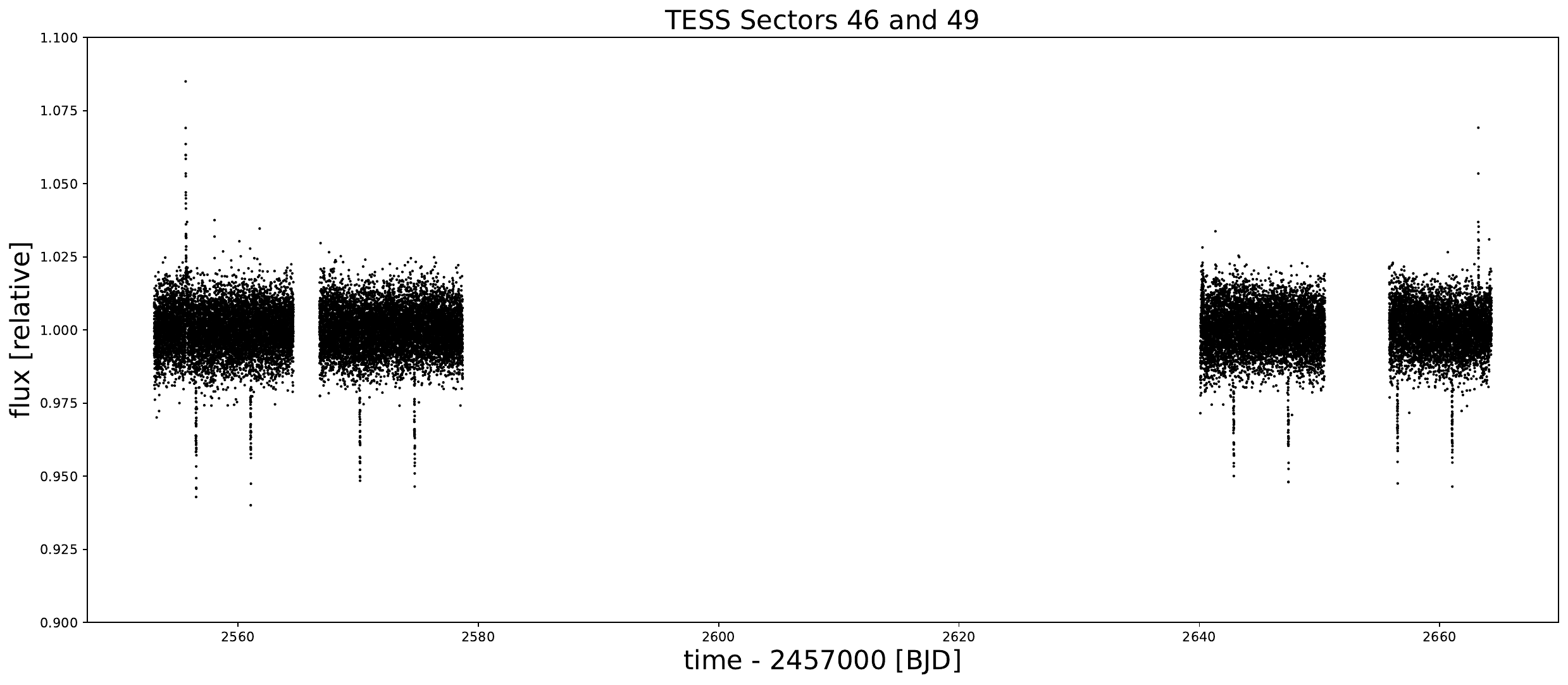}
        \caption{
            Short 2-minute cadence of the TESS Sectors 46 and 49. 
            Both sets of light curves use the PDCSAP flux.
            \href{https://github.com/ssagynbayeva/polka-dotted-stars-toi3884/blob/main/src/tex/notebooks/SSP-TOI3884-analysis.ipynb}{\faGithub}
            }
        \label{fig:toi-3884-pdscap-lc}
    \end{centering}
\end{figure*}

\subsection{Priors for TOI-3884 and TOI-3384 b}

Below, we detail each parameter's prior distribution. We denote the initial values taken from the discovery work of TOI-3884 as 
$\_\rm transit$ taken from \cite{Almenara2022} and the values from the RV observations as $\_\rm RV$ taken from \cite{Libby-Roberts2023}.

The impact parameter $b$ follows a uniform distribution between $-b_{\text{max}}$ and $b_{\text{max}}$, 
where $b_{\text{max}} = R_{\star}/a$ represents the ratio of the stellar radius to the semi-major axis. 
This prior encompasses all possible transit geometries, including grazing transits: $b \sim \mathcal{U}(-b_{\text{max}}, 
b_{\text{max}})$.

The orbital eccentricity $e$ is constrained to physically meaningful values with a uniform prior between 0 (circular orbit) and 1 (parabolic orbit): 
$e \sim \mathcal{U}(0, 1)$.

For the orbital period $P_{\text{orb}}$, we use a tight uniform prior on a scaling factor $f_{P_{\text{orb}}}$ around the known value: 
$f_{P_{\text{orb}}} \sim \mathcal{U}(0.99, 1.01)$, with the actual period derived as $P_{\text{orb}} = P_{\text{orb,transit}} \cdot f_{P_{\text{orb}}}$. 
This allows for small adjustments to the period while maintaining consistency with observed transits.

Similarly, the mid-transit time $t_0$ is parameterized through a scaling factor $f_{t_0} \sim \mathcal{U}(0.9, 1.1)$, 
with the actual time calculated as $t_0 = t_{0,\text{transit}} + P_{\text{orb,transit}} \cdot (f_{t_0} - 1)$. 
This formulation permits the mid-transit time to vary within approximately $\pm10\%$ of the orbital period from the initial estimate.

The planet radius is parameterized in log space with a uniform prior that allows it to vary between half and twice the 
initially estimated value: $\ln(R_p) \sim \mathcal{U}(\ln(R_{p,\text{transit}}/2), \ln(2 \cdot R_{p,\text{transit}}))$.

The orientation of the orbital angular momentum is parameterized through $w_x$ and $w_y$, which follow standard normal 
distributions: $w_x, w_y \sim \mathcal{N}(0, 1)$. 
This provides a uniform distribution over the unit sphere when properly transformed to three dimensions.

The stellar position parameters $\tilde{x}_{\star}$, $\tilde{y}_{\star}$, and
$\tilde{z}_{\star}$ similarly follow standard normal distributions:
$\tilde{x}_{\star}$, $\tilde{y}_{\star} \sim \mathcal{N}(0,1)$, 
$\tilde{z}_{\star}\sim \rm Half\mathcal{N}(0,1)$, allowing for uncertainty in the precise position of the star.

The limb darkening coefficients for a quadratic limb darkening law are given uniform priors: 
$u_1 \sim \mathcal{U}(0, 0.5)$ and $u_2 \sim \mathcal{U}(0, 0.2)$. These ranges are consistent with typical values 
for main-sequence stars.

The stellar rotation period $P_{\text{rot}}$ is parameterized as a multiple of the estimated 
value: $f_{P_{\text{rot}}} \sim \mathcal{U}(0.2, 1.8)$ with $P_{\text{rot}} = f_{P_{\text{rot}}} 
\cdot 10$, allowing the rotation period to vary between 20\% and 180\% of the initial estimate. This translates to the uniform 
distribution from 2 to 18 days.

The stellar mass follows a log-normal distribution derived from density
measurements: $\ln(M_{\star}) \sim \mathcal{N}(\ln(\rho_{\rm RV} /
\rho_{\odot}), \sigma_{\rho_{\rm RV}} / \rho_{\rm RV})$. This prior incorporates observational constraints on the stellar 
density while properly accounting for the uncertainty. Since we fix the radius of the star at 1 $R_\odot$, sampling the 
stellar mass directly translates to sampling the stellar density parameter space, which is the physical quantity 
constrained by our observations. 

For the Gaussian Process (GP) kernel parameters modeling stellar activity, we employ uniform priors: the radius of spots 
$\mathbb{r} \sim \mathcal{U}(10.0, 60.0)$, the contrast $\mathbb{c} \sim \mathcal{U}(0.01, 0.9)$, 
the number of spots $\mathbb{n} \sim \mathcal{U}(0.1, 10.0)$, the mean of the latitude distribution $\mu_\phi \sim \mathcal{U}(0.1, 80.0)$, 
and the variance of the latitude distribution $\sigma^2_\phi \sim \mathcal{U}(0.1, 20.0)$.

All parameters constrained to specific ranges are implemented using logit/inverse-logit transformations with appropriate 
Jacobian adjustments in the log-probability calculations to maintain proper probability distributions.

\subsection{Addressing Multimodality with Parallel Tempering}

The posterior distribution of our model applied to this data set
exhibits significant multimodality, presenting substantial challenges for
conventional sampling methods. This multimodality arises from several sources
inherent to transit modeling. First, the degeneracy between impact parameter
and planet radius creates multiple regions of high posterior probability.
Second, the periodic nature of orbital elements introduces symmetries in the
likelihood landscape, particularly when stellar activity signals interact with
planetary transit signatures. Third, the parameterization of stellar rotation
and activity through our model introduces additional complexity with multiple
viable configurations that can explain the observed data equally well.

Standard Markov Chain Monte Carlo (MCMC) methods such as Ensemble Sampling
(e.g., \texttt{emcee} \cite{emcee}) or Hamiltonian Monte Carlo (HMC;
\cite{Hoffman2011}) struggle in this multimodal landscape. Ensemble samplers can
become trapped in local modes, failing to explore the full posterior
distribution even with large numbers of walkers. This "mode-trapping" problem is
particularly severe in our high-dimensional parameter space, where narrow
connecting pathways between modes create bottlenecks for efficient sampling.
Similarly, HMC methods, while excellent for exploring complicated geometries
within a single mode, typically fail to transition between widely separated
modes.

To overcome these limitations, we utilized Parallel Tempering (PT), also known
as replica exchange MCMC \citep{Earl_2005,ptemcee}. This method
simultaneously runs multiple MCMC chains at different "temperatures," where
temperature controls the degree to which the posterior is flattened. At high
temperatures, the likelihood is down-weighted, allowing chains to move more
freely across the parameter space and easily traverse valleys between modes. At
the lowest temperature ($T=1$), the chain samples from the true posterior of
interest.

The key innovation in PT is the periodic exchange of states between chains at adjacent temperatures. 
These exchanges allow configurations discovered at high temperatures to propagate down to the cold chain, 
effectively enabling jumps between modes that would be extremely rare in standard MCMC. We used the implementation
of PT in \texttt{ptemcee} \citep{ptemcee}. \texttt{ptemcee} handles these exchanges using the Metropolis criterion:

\begin{equation}
    \label{eq:ptemcee}
    P_{\rm exchange} = \min\left(1, \exp\left[\left(\frac{1}{T_i} - \frac{1}{T_j}\right)\left(\ln \mathcal{L}(\theta_j) - \ln \mathcal{L}(\theta_i)\right)\right]\right)
\end{equation}

where $T_i$ and $T_j$ are the temperatures of adjacent chains, and $\theta_i$
and $\theta_j$ are their respective states. Our implementation used a ladder of
8 temperatures.  A key feature
of \texttt{ptemcee} is that it will adjust the temperatures of the ladder as the
simulation progresses to ensure that the rate of exchanges equalizes while
pinning the lowest and highest temperatures to $T=1$ and $T=\infty$.
To ensure convergence, we monitored the integrated autocorrelation time of our sampling chains, confirming that our 
runs exceeded 50 times the maximum autocorrelation length across all parameters.

\subsection{Results}

The full posterior distributions for all model parameters are summarized in Table \ref{tab:ResultsT0i3884}. The table reports the results for planetary orbital inclination $i_p$, eccentricity $e$, the orbital period $P$, the time of the first transit $t_0$, the planet-to-star radius ratio $R_p/R_\star$, the limb-darkening coefficients $u_1$ and $u_2$, the derived radial velocity $v\sin{i}$,  stellar inclination $i_\star$, stellar \textit{projected} obliquity $\lambda_\star$(derived to compare with the literature results; note that the model reports the true obliquity), rotational period of the stellar spin $P_\star$, stellar density $\rho_\star$, radius of the spots $\mathbb{r}$, the contrast of the spots $\mathbb{c}$, number of spots $\mathbb{n}$, the mean $\mu_\phi$ and standard deviation $\sigma_\phi$ of the latitude distribution of spots. Here, we highlight the 
most significant results. 

The spot model strongly favors a high-latitude feature with an angular radius of $\mathbb{r} = 26.77^{+9.7}_{-8.7}$ degrees, 
covering approximately 2.5-10\% of the stellar surface. 

\begin{figure*}[hbt!]
    \centering
    \includegraphics[width=\textwidth]{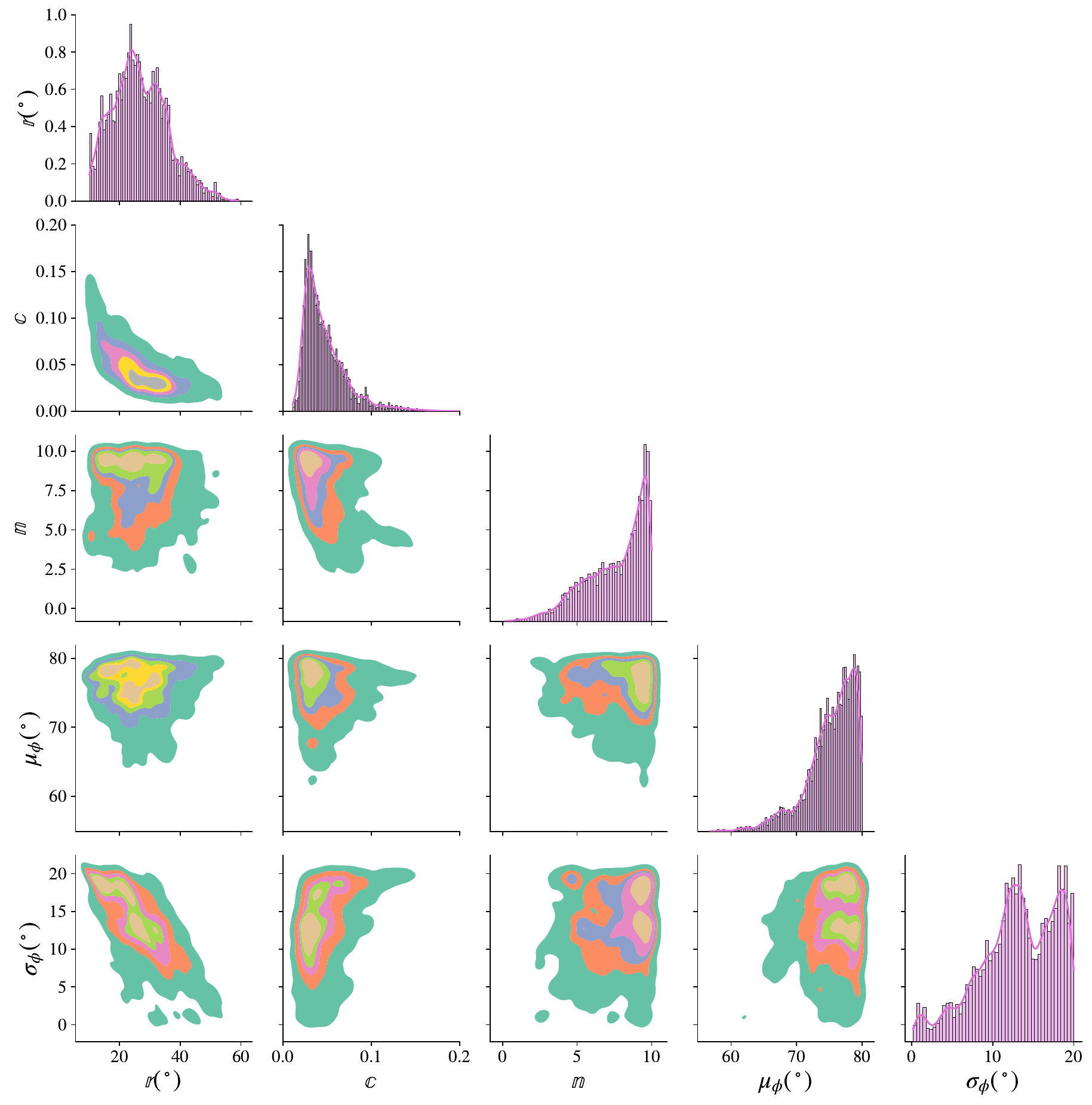}
    \caption{Corner plot showing the posterior distributions and correlations
    for the spot model parameters. The diagonal panels display the marginalized
    posterior distributions for each parameter: the spot size $\mathbb{r}$
    ($^{\circ}$), spot contrast $\mathbb{c}$, number of spots $\mathbb{n}$, spot
    latitude $\mu_{\phi}$ ($^{\circ}$), and the angular variance in latitude
    $\sigma^2_{\phi}$ ($^{\circ}$). The spot latitude is concentrated at high
    values around $75^{\circ}$--$80^{\circ}$, confirming the presence of a
    near-polar spot. The spot contrast is relatively modest with most of the
    probability mass below 0.1. The number of spots tends toward values between
    7.5 and 10, suggesting a relatively small number of spots.
    \href{https://github.com/ssagynbayeva/polka-dotted-stars-toi3884/blob/main/src/tex/notebooks/SSP-TOI3884-analysis.ipynb}{\faGithub}
    \href{https://zenodo.org/records/16647404}{\faDatabase}}
    \label{fig:spot_parameters}
\end{figure*}

Figure \ref{fig:spot_parameters} presents the posterior distributions and correlations of the key parameters in our spot model. 
Our analysis strongly favors a high-latitude spot configuration with the spot center located at $\mu_{\phi} = {75.3}^{+3.7}_{-4.2}$ 
degrees, indicating a near-polar feature. The spot contrast parameter $\mathbb{c} = 0.049^{+0.028}_{-0.013}$ corresponds to a 
moderate temperature difference between the spot and surrounding photosphere, consistent with typical values observed on active M 
dwarfs. The angular variance in longitude $\sigma^2_{\phi}$ shows bimodality, with peaks at approximately $5^{\circ}$ 
and $15^{\circ}$, indicating two distinct possible spot configurations with different latitudinal spreads. 
The close correlation between stellar inclination and spot latitude reflects the geometric constraints imposed by the transit 
observations. Collectively, these parameters paint a picture of TOI-3884 as an active star with significant polar spotting.

\begin{figure*}[hbt!]
    \centering
    \includegraphics[width=\textwidth]{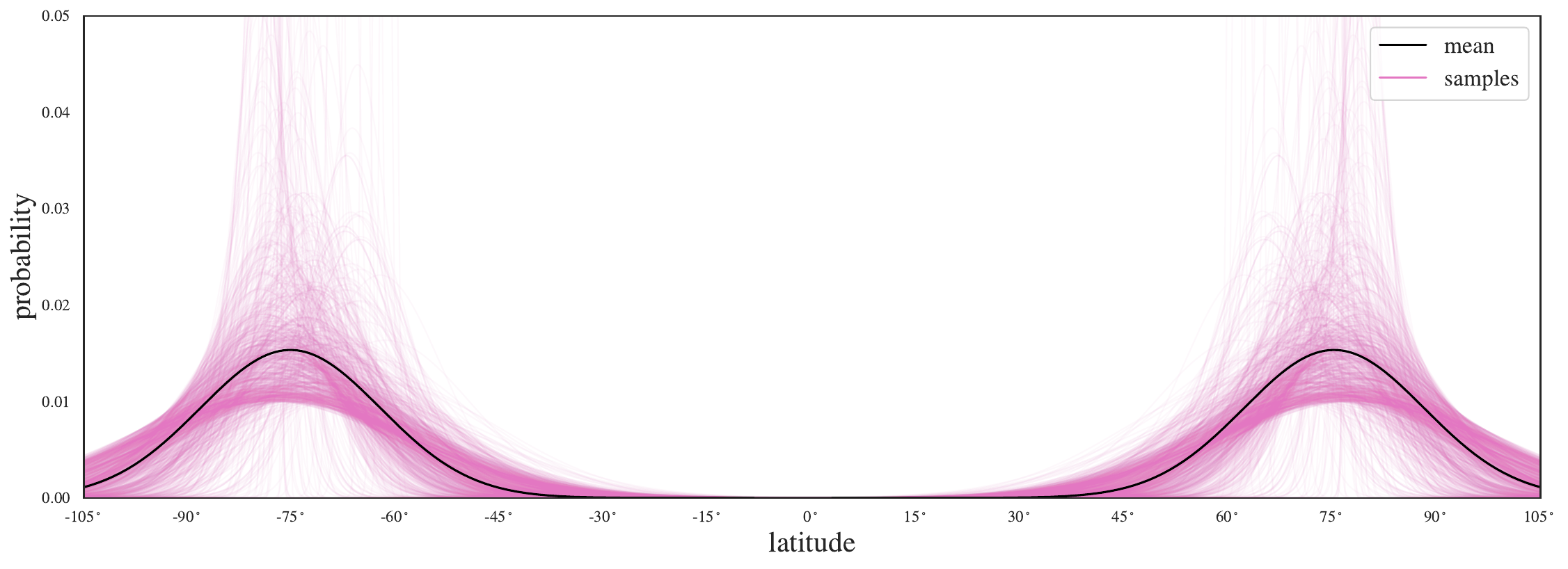}
    \caption{Posterior probability distribution of spot latitudes for TOI-3884.
    The black line shows the mean distribution, while the pink lines represent
    individual posterior samples from our MCMC analysis. The distribution peaks
    at high latitudes ($\pm75^{\circ}$) with minimal probability near the
    equator, indicating a strong preference for near-polar spots. The symmetry
    across the equator reflects the inherent degeneracy in determining the
    hemisphere of spot locations from transit data alone.
    \href{https://github.com/ssagynbayeva/polka-dotted-stars-toi3884/blob/main/src/tex/notebooks/SSP-TOI3884-analysis.ipynb}{\faGithub}
    \href{https://zenodo.org/records/16647404}{\faDatabase}}
    \label{fig:spot_latitudes}
\end{figure*}

Our analysis of the spot latitude distribution is illustrated in Figure \ref{fig:spot_latitudes}. This figure shows the posterior 
probability distribution of spot latitudes derived from the model. The black line represents the mean distribution, 
while the individual pink lines show a subset of 1000 posterior samples, highlighting the inherent variability in our constraints. 
The distribution exhibits two prominent peaks centered at approximately $-75^{\circ}$ and $+75^{\circ}$ latitude, 
with negligible probability near the equator.

While we cannot definitively determine whether the spot is in the northern or
southern hemisphere based on the transit data alone, the presence of either
configuration would lead to similar physical interpretations regarding the
underlying stellar dynamo processes. This latitude distribution, combined with
the other spot parameters discussed previously, presents compelling evidence for
strong, organized magnetic fields concentrated at the rotational poles of
TOI-3884 \citep{Libby-Roberts2023}. 

The stellar inclination angle was found to be $i_\star = {34.8}^{+5.62}_{-6.17}$ degrees, indicating that the stellar rotation 
axis is considerably inclined from the line of sight. More notably, the sky-projected spin-orbit angle (stellar obliquity) was 
determined to be $\psi_\star = {80.37}^{+49.6}_{-27.5}$ degrees.
The stellar rotation period was constrained to $P_\text{rot} = 9.07^{+0.45}_{-0.51}$ days, which places TOI-3884 among the 
more rapidly rotating M dwarfs, consistent with its observed activity level and estimated age.

We find a planetary radius of $R_p = 0.188 \pm 0.003$ $R_\star$ and an orbital period of $P_\text{orb} =
4.5446^{+2.7\times10^{-5}}_{-2.4\times10^{-5}}$ days. 

\begin{figure*}[hbt!]
    \centering
    \includegraphics[width=\textwidth]{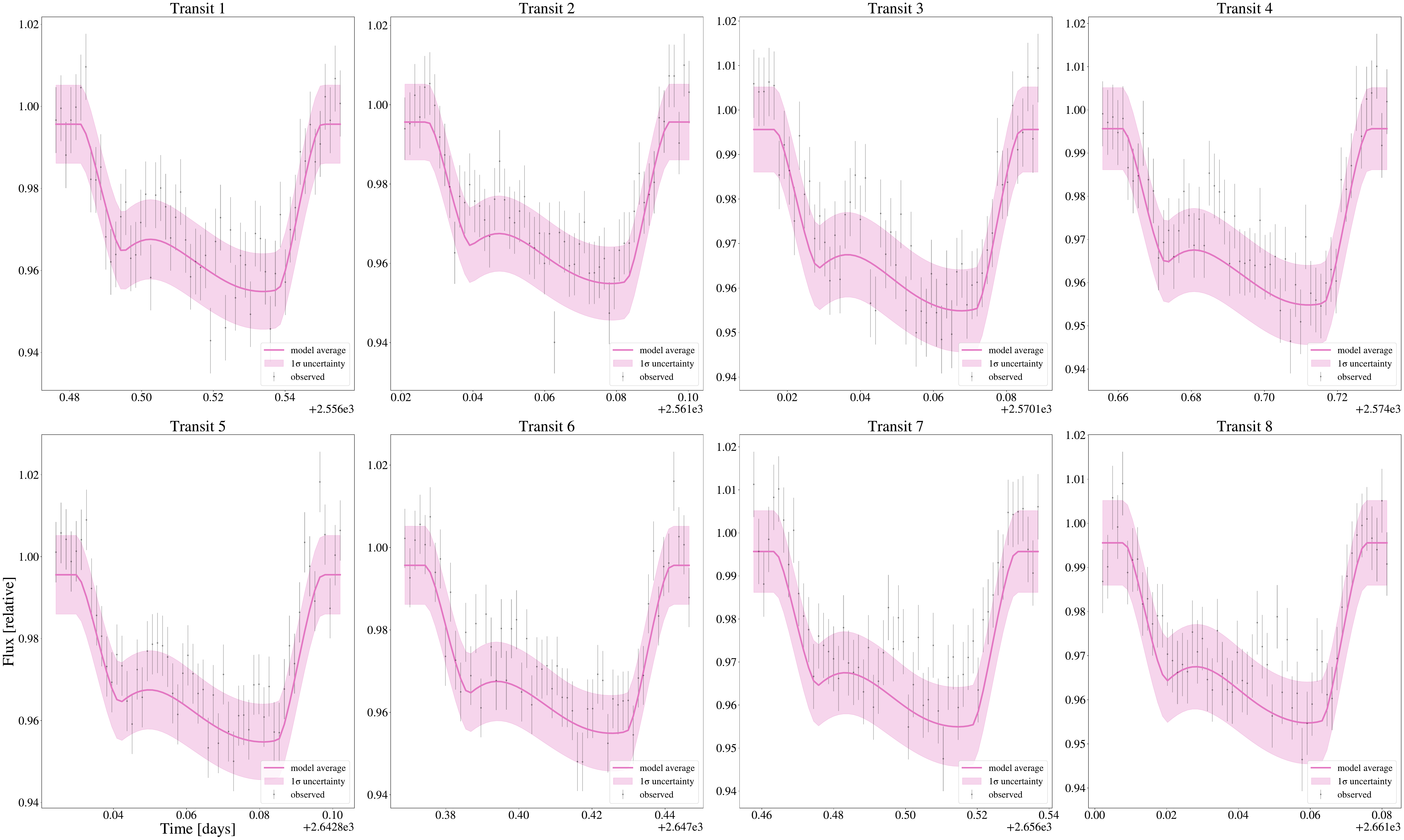}
    \caption{Light curves for eight transit events of TOI-3884b. Each panel shows the observed photometric data from TESS (gray points 
    with error bars) and the results of our MCMC modeling. The solid pink lines represent the model average from 100 posterior 
    samples, while the light pink shaded regions indicate the 1$\sigma$ uncertainty bounds. The x-axis displays time in days 
    relative to the mid-transit point (shown as offsets at the bottom of each panel), and the y-axis shows the normalized flux. 
    The slight asymmetries and variations between transit epochs are consistent with the influence of a high-latitude stellar spot. 
    Our model successfully reproduces the observed transit profiles across all epochs, supporting our inferred spot and stellar 
    orientation parameters.
    \href{https://github.com/ssagynbayeva/polka-dotted-stars-toi3884/blob/main/src/tex/notebooks/SSP-TOI3884-analysis.ipynb}{\faGithub}
    \href{https://zenodo.org/records/16647404}{\faDatabase}}
    \label{fig:toi3884-transits}
\end{figure*}

\begin{table}[]
    \vspace{0.5cm}
    \centering
    \caption{The free parameters for TOI-3884 data in Section \ref{sec:toi3884}, their true values, and their priors.}
    \begin{tabular}{lll}
    \hline
    Parameter                                 & Inferred value    & \cite{Libby-Roberts2023} \\ \hline\hline
    $i_p (^\circ)$                            & $90.17^{+0.52}_{-0.78}$                            & $89.81^{+0.13}_{-0.18}$                    \\
    $e$                                       & $0.14^{+0.19}_{-0.07}$                             & $0.06^{+0.06}_{-0.04}$                     \\
    $P (\rm days)$                            & $4.5446^{+2.7\times10^{-5}}_{-2.4\times10^{-5}}$                                                                                              & $4.5445828\pm 0.0000098$                   \\
    $t_0 (\rm days)$                          & $2556.51\pm 4\times 10^{-4}$                       &    -                 \\
    $R_p / R_\star$                           & $0.188 \pm 3\times10^{-3}$                         &  $0.197\pm 0.002$                   \\ 
    $u_1$                                     & $0.14^{+0.15}_{-0.07}$                             &  -                   \\ 
    $u_2$                                     & $0.09 \pm 0.06$                                    &  -                   \\ 
    $v\sin i (\rm kms^{-1})$                  & $1.69^{+0.11}_{-0.09}$                             & $3.59\pm 0.92$                    \\ \hline
    $i_\star (^\circ)$                        & $34.8^{+5.62}_{-6.17}$                             &  $25\pm 5$                    \\
    $\lambda_\star (^\circ)$                  & $80.58^{+48.3}_{-27.5}$                            &  $75\pm 10$                   \\
    $P_\star (\rm days)$                      & $9.07^{+0.45}_{-0.51}$                             & $<4.22\pm 1.09$                    \\ 
    $\rho_\star (\rm gcm^{-3})$               & $15.18^{+1.98}_{-1.75}$                            & $15.26\pm 2.04$                    \\ \hline
    $\mathbb{r} (^\circ)$                     & $26.77^{+9.7}_{-8.7}$                              &  -                   \\
    $\mathbb{c}$                              & $0.049^{+0.028}_{-0.013}$                          &  -                   \\
    $\mathbb{n}$                              & $7.58^{+1.39}_{-2.85}$                             &  -                   \\
    $\mu_\phi (^\circ)$                       & $75.24^{+2.67}_{-4.11}$                            &  -                   \\
    $\sigma_\phi (^\circ)$                  & $12.99^{+4.91}_{-5.08}$                              &  -                   \\ \hline
    \label{tab:ResultsT0i3884}
    \end{tabular}
\end{table}
Figure \ref{fig:toi3884-transits} presents the transit light curves of TOI-3884
from eight separate transit events. Each panel shows the photometric data (gray
points with error bars) along with the results from our model. The solid pink
lines represent the model average from 100 posterior samples, while the light
pink shaded regions indicate the $1\sigma$ uncertainty bounds of our model. The
x-axis shows time in days relative to the mid-transit point (indicated by the
offset values), and the y-axis displays the normalized flux.

\begin{figure}[hbt!]
    \centering
    \includegraphics[width=0.46\textwidth]{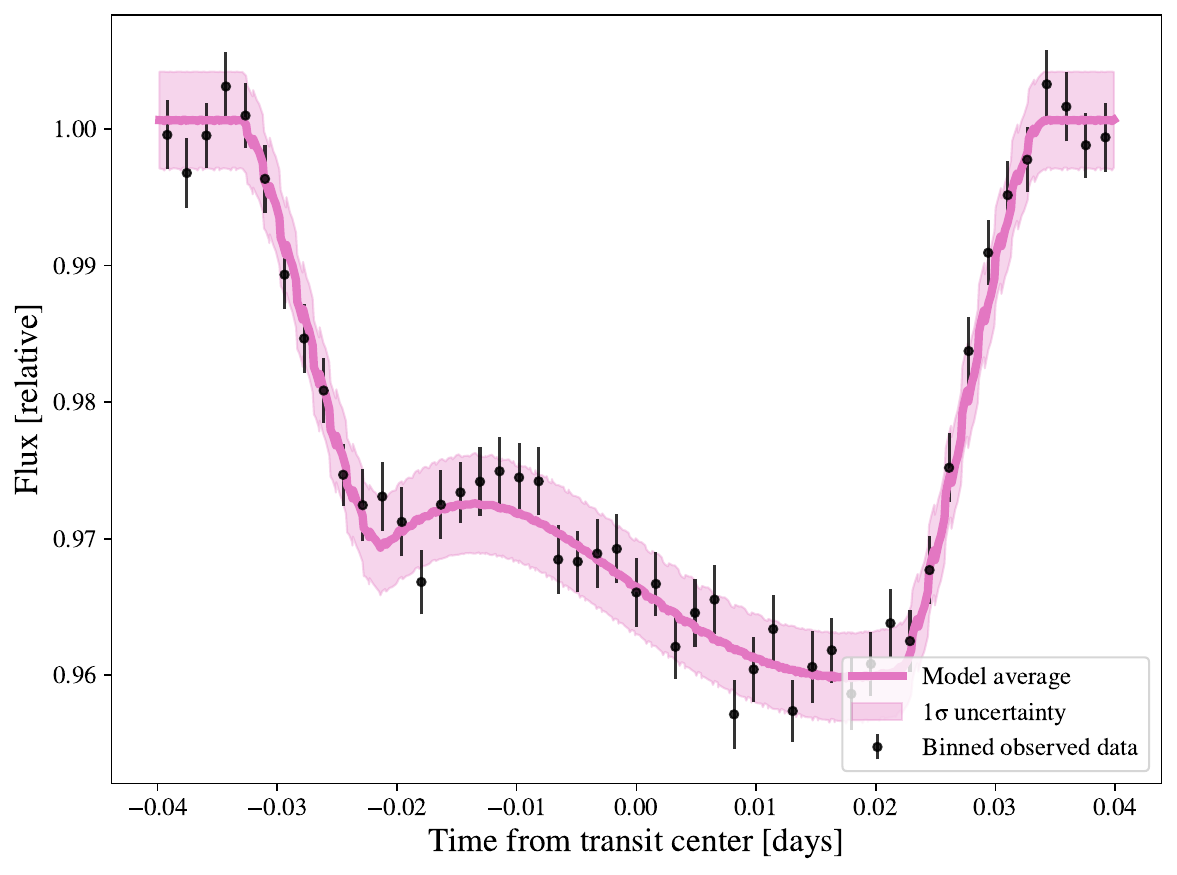}
    \caption{Light curves binned and averaged across the eight transit events of TOI-3884b. The plot shows the observed photometric data from TESS (black points 
    with error bars) and the results of our MCMC modeling averaged across transits and binned in time. The solid pink lines represent the model average from 100 posterior 
    samples, while the light pink shaded regions indicate the 1$\sigma$ uncertainty bounds. The x-axis displays time in days 
    relative to the mid-transit point, and the y-axis shows the normalized flux. 
    The similarity of the binned average transit to the individual transits shown in Figure \ref{fig:toi3884-transits} is consistent with a static polar spot.
    \href{https://github.com/ssagynbayeva/polka-dotted-stars-toi3884/blob/main/src/tex/notebooks/SSP-TOI3884-analysis.ipynb}{\faGithub}
    \href{https://zenodo.org/records/16647404}{\faDatabase}}
    \label{fig:toi3884-transits-avg}
\end{figure}

These transit light curves reveal several notable features. The slight asymmetry
visible in the transit profiles provides strong evidence for the stellar surface
inhomogeneities due to spot-crossings. In particular, the subtle variations
between different transit epochs (particularly visible in transits 3, 4, and 8)
are consistent with the high-latitude spot configuration we've identified.
This implies that the spot is not a uniform polar cap or that
it evolves in time. Our model successfully captures the essential
features of all eight transit events, including the ingress and egress shapes as
well as the in-transit variations. The quality of the fit across multiple epochs
demonstrates the robustness of our spot model and stellar orientation
constraints. The consistency between the observed data and our model supports
our conclusion that TOI-3884 hosts a significant near-polar spot that influences
the transit morphology across multiple planetary orbits. The consistent shape of the model across all transits indicates that the spot is likely static rather than evolving over time. Figure \ref{fig:toi3884-transits-avg} demonstrates this: the binned average across transits matches our model's binned average, and both share the same shape as the binned data average. While the model does not \textit{require} the spot to be static, this is what the model \textit{finds}.


\begin{figure*}[hbt!]
    \centering
    \includegraphics[width=\textwidth]{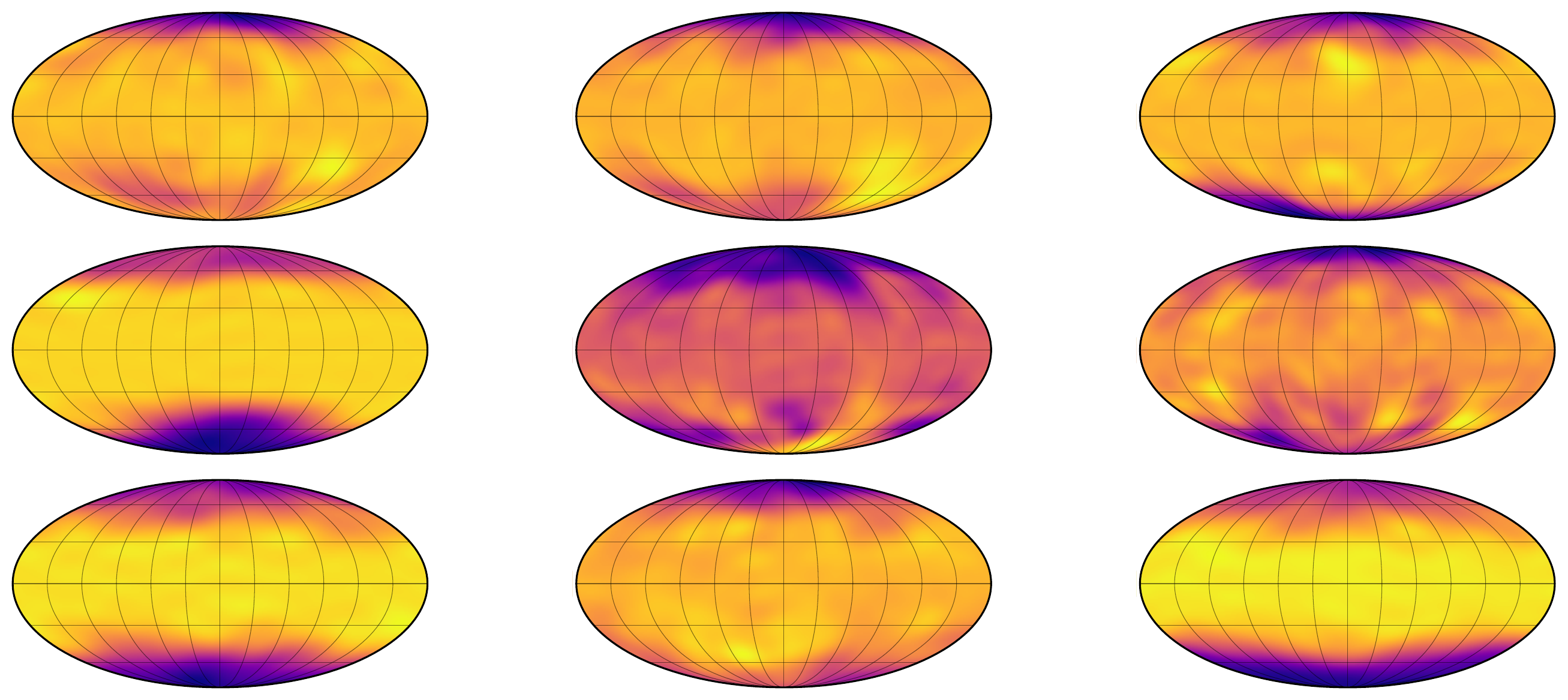}
    \caption{Mollweide projections of the stellar surface of TOI-3884 showing spot distributions from nine representative 
    posterior samples. The majority of samples exhibit prominent high-latitude spot concentrations, with some variation in 
    the exact morphology and distribution. This consistent pattern of near-polar spots across different posterior samples 
    provides strong support for our inferred high-latitude magnetic activity. 
    These surface configurations produce the transit light curve modulations observed in Figure~\ref{fig:toi3884-transits}.
    \href{https://github.com/ssagynbayeva/polka-dotted-stars-toi3884/blob/main/src/tex/notebooks/SSP-TOI3884-analysis.ipynb}{\faGithub}
    \href{https://zenodo.org/records/16647404}{\faDatabase}}
    \label{fig:toi3884-moll}
\end{figure*}

Figure \ref{fig:toi3884-moll} presents a visualization of the stellar surface of TOI-3884 with the modeled spot 
distributions from nine different posterior samples. Each panel displays a Mollweide projection of the entire stellar surface, 
with the color scale indicating intensity variations. The dark purple regions represent cooler areas corresponding to stellar 
spots, while the yellow-orange regions depict the warmer, unspotted photosphere.
These visualizations clearly demonstrate the high-latitude spot configurations. Across the different posterior samples, 
we consistently observe spot concentrations near the poles, though with some variation in the exact morphology and distribution. 
The variations between panels illustrate the range of possible spot configurations that are consistent with our transit 
observations. Despite these differences, the predominance of high-latitude spots is a robust feature across all samples, 
supporting the conclusion that TOI-3884 hosts significant near-polar magnetic activity. 
The consistency of these high-latitude features across different posterior samples demonstrates that this result is 
not sensitive to the specific parameter values within our model's posterior distribution.
These surface maps provide a direct visualization of the spot configuration that gives rise to the transit light curve 
asymmetries shown in Figure \ref{fig:toi3884-transits}, illustrating the three-dimensional 
geometry of the TOI-3884 system.

\begin{figure*}[hbt!]
    \centering
    \includegraphics[width=\textwidth]{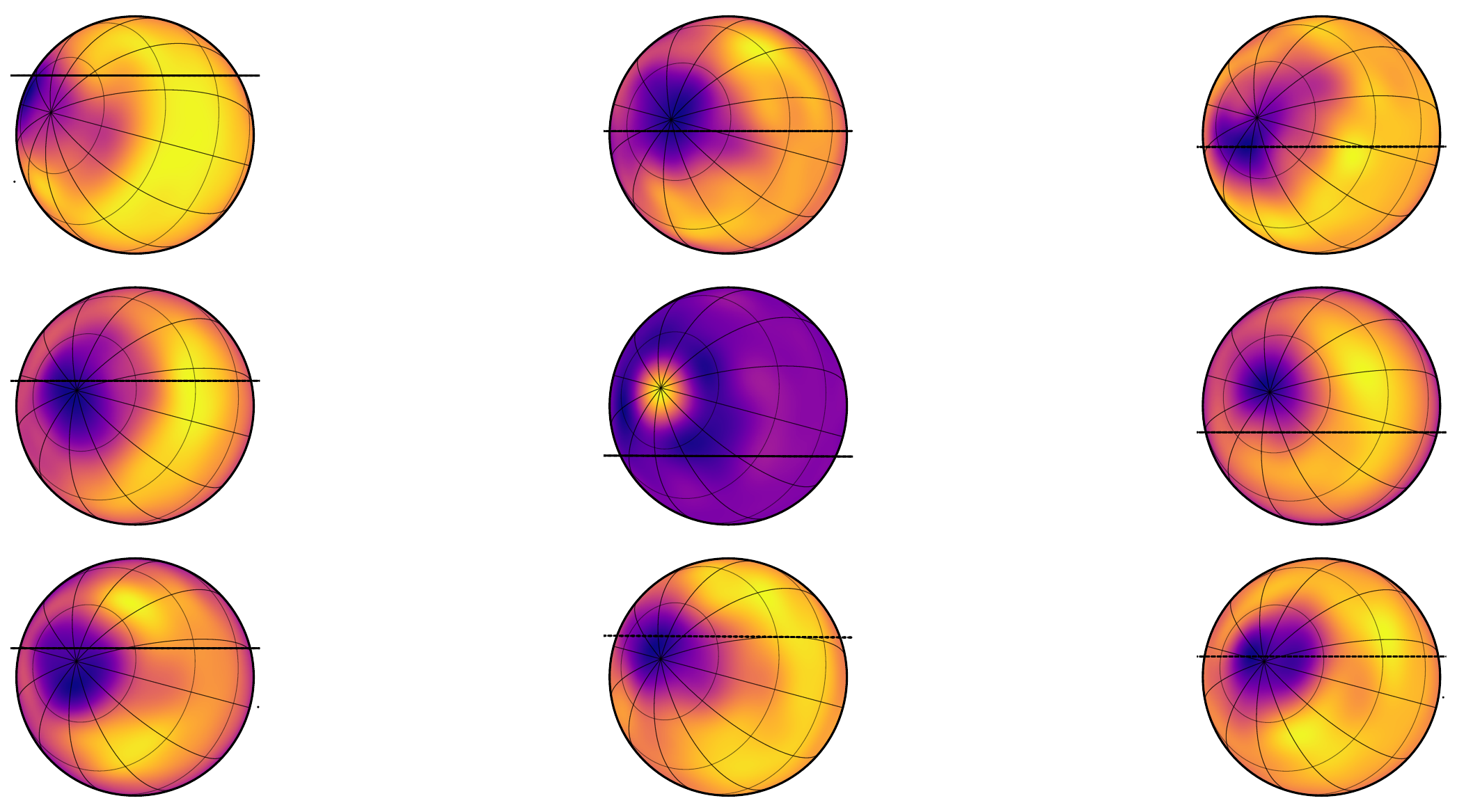}
    \caption{Observer's view of TOI-3884 during transit events from nine representative posterior samples. 
    Each panel shows the stellar disk as viewed from Earth, with dark purple regions indicating cooler spots and 
    yellow-orange areas representing the warmer photosphere. Black dots trace the path of TOI-3884b across the stellar disk 
    during transit. The consistent presence of high-latitude spots intersecting or near the planet's trajectory explains the 
    asymmetries observed in the transit light curves. This visualization demonstrates how the combination of stellar inclination 
    ($i_\star = {34.8}^{+5.62}_{-6.17}$ degrees), spot distribution, and planetary orbital geometry produces the distinctive
    transit signatures in the TESS data.
    \href{https://github.com/ssagynbayeva/polka-dotted-stars-toi3884/blob/main/src/tex/notebooks/SSP-TOI3884-analysis.ipynb}{\faGithub}
    \href{https://zenodo.org/records/16647404}{\faDatabase}}
    \label{fig:toi3884-traj}
\end{figure*}

Similarly, Figure \ref{fig:toi3884-traj} shows nine representative views from our posterior samples. Each panel displays the stellar disk as it 
appears from Earth, with the temperature variation across the surface indicated by the color gradient. The black dots trace 
the path of TOI-3884b across the stellar disk during transit, clearly illustrating how the planet's trajectory intersects 
with surface features. As the planet crosses regions of different brightness on the stellar surface, it blocks varying amounts 
of light, creating the subtle but distinctive signatures we observed in the data. The consistent presence of near-polar spots 
across different posterior samples, combined with the specific transit geometry, provides strong evidence for our inferred 
low stellar inclination and high stellar obliquity. These results illustrate the three-dimensional relationship 
between the planet's orbital plane and the star's rotation axis, supporting the significant spin-orbit misalignment in the 
TOI-3884 system.

\section{Discussion}
\label{sec:discussion}
Our analysis demonstrates that transit light curves can provide meaningful constraints on stellar surface features and 
three-dimensional geometry, even when using single-band photometric data. The model successfully recovers a consistent spot 
distribution across multiple posterior samples (see Section \ref{sec:experiment1}), characterized predominantly by high-latitude 
features in the case of TOI-3884 (see Section \ref{sec:toi3884}). 
This result is robust across our MCMC exploration and explains the observed transit light curve morphologies. 
However, there are several important considerations regarding the model's limitations and strengths.

\subsection{Parameter Degeneracies and Uncertainty}

While our model accurately reproduces the observed transit light curves, the posterior distributions for several key 
parameters remain relatively wide. This breadth reflects inherent degeneracies in transit modeling that persist even 
with our approach described in Section \ref{sec:model}. For instance, the stellar inclination angle 
($i_\star$) and sky-projected spin-orbit angle 
($\psi_\star$) have substantial uncertainties. These uncertainties stem 
from multiple parameter combinations that can produce similar transit signatures, particularly regarding the precise spot 
locations and their contrast with the surrounding photosphere.

The degeneracy between northern and southern hemisphere spot locations is particularly evident. As shown in Figure 
\ref{fig:spot_latitudes} and also discussed in Section \ref{sec:model}, we cannot definitively distinguish 
between spots located at positive or negative latitudes using 
transit data alone. This ambiguity is fundamental to the transit method, as the planet's path across the stellar disk 
cannot discriminate between spots above or below its trajectory when they are symmetric about the transit chord.

Despite these degeneracies, our flexible spot model recovers a consistent overall picture of TOI-3884's surface features. 
The near-polar spot configuration emerges as a robust result regardless of which specific parameter combination within our 
posterior is considered. This consistency suggests that while individual parameter values may be uncertain, the general 
surface feature distribution is well-constrained by the data.

\subsection{Recovery of 3D Information from Photometry Alone}

A significant achievement of our approach is the recovery of three-dimensional information about the star-planet system using 
only single-band photometry. Previous studies typically require complementary data sources such as Doppler tomography or 
spectropolarimetric observations to constrain the three-dimensional geometry of transiting systems. 
Our work demonstrates that with appropriate modeling, transit light curves themselves contain significant information about 
the stellar rotation axis and surface feature distribution.

Our derived stellar projected rotational velocity ($v\sin i_\star = {5.2}^{+0.7}_{-0.8}$ km/s) is in good agreement 
with spectroscopically measured values for TOI-3884 in \cite{Libby-Roberts2023}, despite being derived s
olely from photometric data. This agreement provides strong independent validation of our geometric model 
and suggests that photometric analyses can serve as useful complements to spectroscopic methods for determining stellar 
rotational properties.

Our results highlight important limitations in previous spot modeling approaches. For instance, 
the models employed before assumed either a single spot or a fixed number of discrete spots 
\citep{Morris2018,Oshagh2013,Libby-Roberts2023,Beky2014}, while \cite{Huber2010} used longitudinal bands. 
While computationally efficient, 
such strict assumptions artificially reduce uncertainty about the stellar orientation and spot configuration. 
By imposing strong priors on the number and morphology of spots, these models may underestimate the true parameter 
uncertainties and potentially miss important surface features.
Our Gaussian process-based approach allows for more flexible spot distributions with varying complexity, 
providing a more realistic assessment of the constraints that can be derived from transit data. 
The wider posterior distributions in our analysis more accurately reflect the genuine uncertainties inherent in 
inferring stellar surface features from transit light curves alone. 

\begin{figure*}[hbt!]
    \centering
    \includegraphics[width=\textwidth]{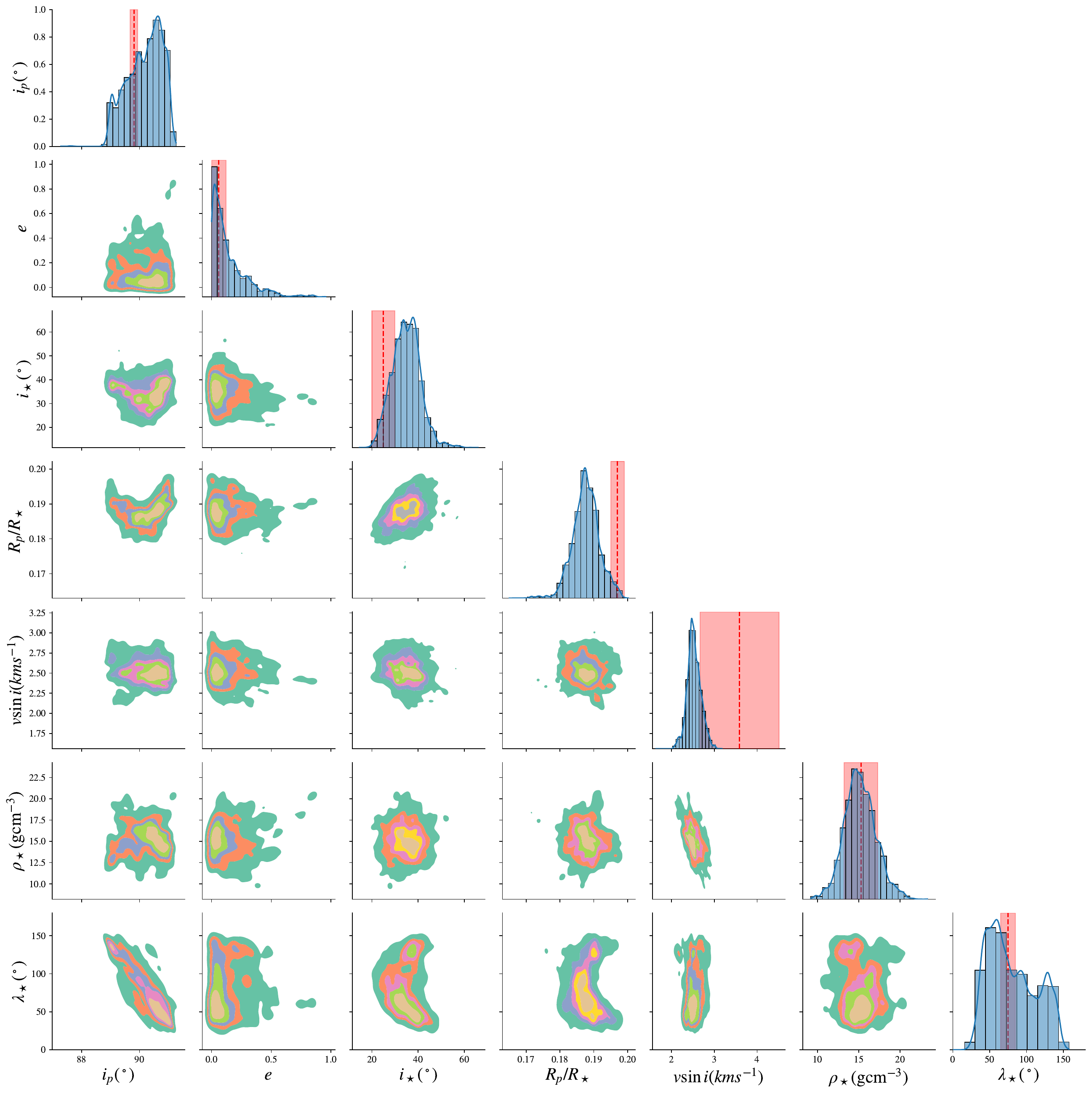}
    \caption{The corner plot for planet's inclination ($i_p$), orbital
    eccentricity ($e$), stellar inclination ($i_\star$), sky-projected
    spin-orbit angle ($\lambda_\star$), planet-to-star radius ratio
    ($R_p/R_\star$), projected stellar rotational velocity ($v\sin i$), and
    stellar density ($\rho_\star$). The red shaded regions and vertical dashed
    lines indicate independent 1-$\sigma$ constraints from
    radial velocity measurements reported by \cite{Libby-Roberts2023}.
    \href{https://github.com/ssagynbayeva/polka-dotted-stars-toi3884/blob/main/src/tex/notebooks/SSP-TOI3884-analysis.ipynb}{\faGithub}
    \href{https://zenodo.org/records/16647404}{\faDatabase}}
    \label{fig:toi3884-compare}
\end{figure*}

Figure \ref{fig:toi3884-compare} presents the posterior distributions and correlations for selected parameters in our model of 
TOI-3884. The parameters shown include the planet's inclination ($i_p$), orbital eccentricity ($e$), 
stellar inclination ($i_\star$), sky-projected spin-orbit angle ($\lambda_\star$), planet-to-star radius ratio ($R_p/R_\star$), 
projected stellar rotational velocity ($v\sin i$), and stellar density ($\rho_\star$).
The red shaded regions and vertical dashed lines indicate independent constraints from radial velocity measurements reported by 
\cite{Libby-Roberts2023}. These RV-derived constraints provide important independent validation for our photometry-based model. 
Our photometrically derived $v\sin i$ value of approximately 2.5 km/s aligns within 1$\sigma$-well with the spectroscopic 
measurement, 
despite being derived solely from transit light curves. This agreement strongly reinforces the power of our approach to 
extract detailed stellar and orbital parameters using only photometric data.
We show that with single-band photometry alone, we can constrain parameters traditionally measured through spectroscopic techniques. 
The ability to recover $v\sin i$ and place meaningful constraints on the spin-orbit alignment without recourse to 
spectroscopy represents a significant methodological advancement, potentially enabling detailed characterization of 
systems for which high-resolution spectroscopy is challenging.

\subsection{Evolving Surface}
The evolving surface model presented in Sections \ref{sec:evolmodel} and \ref{sec:experiment2} demonstrates a promising 
approach for tracking stellar surface feature changes over time, and definitely needs to be imroved further. 
However, it is important to note that there are multiple strategies for addressing time-varying stellar surfaces in transit 
modeling. While our preliminary evolving model incorporates time-dependent surface features through a linear interpolation, 
we found that a simpler approach can also be effective for many applications.

For TOI-3884, we adopted a pragmatic solution by segmenting the light curve into shorter time intervals and applying 
our static spot model (Section \ref{sec:experiment1}) to each segment independently. This approach is justified when 
the segments are shorter than the characteristic evolution timescale of stellar surface features. Based on empirical studies 
of active stars, we chose a segment duration of approximately half the stellar rotation period, 
which ensures that surface evolution within each segment is minimal while still providing sufficient data for robust model fitting.
This segmented approach offers several advantages. First, it avoids the additional complexity and computational expense of a 
fully time-dependent model while still capturing the essential time evolution of the stellar surface. Second, it allows for 
direct comparison between epochs, enabling the identification of persistent features (like the high-latitude spot we detected) 
versus transient ones. Third, it provides a more straightforward statistical interpretation since each segment's model is 
independent of the others.
The results from our segmented analysis of TOI-3884 reveal that while some surface features evolve over the observational 
baseline, the high-latitude spot configuration remains relatively stable. 

For future work, one should compare the results from this segmented approach with fully time-dependent models for determining 
the optimal modeling strategy for different types of stars and observational scenarios, which is beyond the scope of this paper.
Stars with rapid surface evolution or observations spanning many rotation periods might benefit more from the fully 
time-dependent approach, while observations covering only a few rotation periods may be adequately modeled with the 
segmented approach.
Additionally, the segmented approach provides a practical starting point for characterizing surface evolution, 
from which more sophisticated time-dependent models can be developed. The evolving surface model presented 
here represents an initial step toward such comprehensive modeling, which will become increasingly important as 
higher-precision photometry becomes available from current and future missions.

\subsection{Small Spots} \label{sec:smallspots}
An important limitation of our current model concerns the spatial resolution of stellar surface features. 
Our implementation uses spherical harmonics with a maximum degree of $l_{\rm max}=15$, which provides sufficient 
resolution to capture large-scale spot structures like the high-latitude features we identified on TOI-3884, or the spots of size 
$\sim 10^\circ$ like in Section \ref{sec:experiment1}. 
However, this resolution is inadequate for modeling the small-scale spot distributions characteristic of the Sun.
The Sun exhibits numerous small spots with typical sizes of just a few degrees across the stellar surface, often appearing 
in complex active regions. Such small-scale features can significantly impact transit depth measurements while being difficult 
to resolve with current modeling approaches. \cite{Morrisradius} showed that using ingress/egress durations rather than 
transit depths can provide more accurate radius measurements when small spots are present on the stellar surface.
Still, accurately representing such fine-scale structures would require substantially higher spherical 
harmonic degrees, potentially $l_{\rm max} > 100$ \citep{Solanki2003}. The computational challenge of such high-resolution modeling is very significant. 
The size of the covariance matrix in our approach scales as $(l_{\rm max}+1)^2 \times (l_{\rm max}+1)^2$, meaning that 
doubling the maximum degree increases the memory requirements by a factor of 16 and the computational time by even more.

While the Sun does not exhibit spots as large as 10 degrees, young stars may present extensive active regions. It is important 
to clarify, however, that these large active regions differ fundamentally from sunspots as traditionally defined bipolar 
magnetic regions (BMRs). Individual BMRs of 10-degree scale are physically improbable based on our understanding of 
stellar magnetohydrodynamics. Nevertheless, clusters of BMRs that collectively span approximately 10 degrees are 
plausible. This distinction between individual large spots and clustered active regions should be acknowledged when 
interpreting results derived from spherical harmonic-based brightness reconstructions of stellar surfaces.

For TOI-3884, the $l_{\text{max}}=15$ resolution is justified by what we know about active M dwarfs, which tend to form 
larger spot structures than G-type stars like the Sun. Zeeman-Doppler imaging and spectropolarimetric studies of M dwarfs 
generally reveal simpler, larger-scale magnetic field configurations, particularly for fully convective stars. 
However, it remains unclear whether these stars truly lack small-scale features or if current observational techniques 
simply cannot resolve them.

Additionally, we note that the effective resolution achievable from transit light curves is inherently limited by the transit 
chord width and photometric precision. Even with perfect computational resources, the information content of the transit 
signal places fundamental constraints on the recoverable spot size distribution. Multi-wavelength transit observations 
would provide additional constraints that could help justify the increased computational expense of higher-resolution 
models \citep{Gastine2013,Strassmeier2009}.

Given these considerations, our current $l_{\text{max}}=15$ model represents a practical compromise between resolving 
power and computational feasibility. It captures the dominant features affecting the transit light curves while remaining 
tractable for Bayesian inference with current computational resources. The approach can be scaled to higher resolutions as 
needed when studying stars where smaller spot structures significantly impact the observations.

While we reference active M dwarfs as 
potential comparative cases for more magnetically active stars, they require careful consideration.
The \texttt{StarryProcess} framework is fundamentally structured around the concept of active latitudes, 
a characteristic pattern observed in stars with solar-like dynamo mechanisms. The dynamo mechanisms in M dwarfs 
remain not completely understood, and there is no conclusive observational evidence that these stars exhibit the 
active latitude patterns that underpin our model. Indeed, theoretical predictions for $\alpha^2$-dynamos often suggest more 
uniform spot distributions across stellar surfaces, with activity potentially concentrating at polar regions in rapidly 
rotating stars \citep{Yadav2015,Gastine2013,Chabrier2006}. 

\subsection{Implications for M Dwarf Magnetic Activity}

The high-latitude spot configuration we detect aligns with theoretical expectations for rapidly rotating low-mass stars like 
TOI-3884. Magnetohydrodynamic simulations predict that Coriolis forces in such stars should preferentially channel 
magnetic flux emergence toward the rotational poles \citep{Yadav2015,Almenara2022}. Our results provide observational support for these theoretical predictions, 
demonstrating that transit modeling can be a valuable tool for studying stellar magnetic activity patterns.
The spot distribution on TOI-3884 is consistent with studies of other active M dwarfs, where polar spots have been observed 
to persist for multiple rotation periods or even longer \citep{Barnes2017,Morin2008,Strassmeier2009}. It should be noted, though,
that for a better understanding of the magnetic dynamo on M dwarfs, the extensive study on the whole population of different
stars of different spectral types is necessary. 
For example, \cite{Morin2008} showed spectropolarimetric observations of five mid-M dwarf stars 
(including GJ 65A/B binary components), and showed that these fully-convective or near-fully-convective stars possess 
mainly strong, axisymmetric, poloidal magnetic fields with minimal differential rotation, contrasting with magnetic 
topologies observed in more massive G and K stars. However,
\cite{Barnes2017} showed Doppler imaging observations of three fully convective M dwarf stars, revealing surprisingly 
different starspot distributions between the near-identical binary components GJ 65A and GJ 65B, with GJ 65A showing 
high-latitude circumpolar spots similar to GJ 791.2A, while GJ 65B displays larger spots concentrated at intermediate 
latitudes, suggesting different dynamo mechanisms may be operating despite their similar masses and rotation rates.

Studies focused on flares, which represent extremely 
localized magnetic phenomena, provide complementary evidence for this polar concentration of magnetic activity. In particular,
\cite{Ilin2021} present an analysis of long-duration flares on fully convective stars observed with TESS. 
They demonstrate that all four detected giant flares occurred at remarkably high latitudes, far higher than typical solar 
flare latitudes. Additionally, \cite{Ilin2023} developed an ensemble approach to determine flare latitudes which indicates 
that magnetic field emergence in active M dwarfs tends to occur preferentially at high latitudes, particularly for 
rapidly rotating stars. 

The combination of a moderately inclined stellar rotation axis and significant high-latitude spots also helps explain why 
previous analyses of TOI-3884 detected relatively low photometric variability despite evidence for active chromospheric emission. 
High-latitude spots maintain a relatively constant projected area as the star rotates when viewed at moderate inclinations, 
resulting in reduced rotational modulation while still affecting transit profiles when intersected by the planet's path.

Finally, for analyzing JWST observations that typically cannot span a complete stellar rotation period,
the transit data alone could also yield valuable information about the starspot distribution under certain conditions. 
This approach would require making well-justified assumptions about the stellar obliquity and inclination, as these 
parameters critically influence which regions of the stellar surface are sampled during transit events. This highlights both 
a limitation and an opportunity - while complete surface maps need full phase coverage, 
even the spatially and temporally constrained information from high-precision transit observations can significantly 
advance our understanding of stellar activity patterns on TOI-3884 and similar active host stars.

\subsection{Computational Limitations}
The model's computational demands scale substantially with dataset size and resolution, at leading order proportional to $\ell_\mathrm{max}^4 * N_t$; for the examples in this paper on modern hardware, we  
required approximately one week of single-core processing time for TESS-scale datasets and several weeks for 
larger Kepler datasets. While this computational cost reflects the inherent complexity of jointly modeling high-dimensional 
parameter spaces across multiple hierarchical levels, we recognize that such processing times may limit immediate 
applicability to the largest available datasets. Despite these computational challenges, we argue that the method's 
ability to provide robust uncertainty quantification and physically motivated parameter estimates justifies the 
increased computational investment, particularly for detailed studies of individual systems.

\subsection{Future Work}
The methodology presented in this work opens several promising avenues for future research and applications. The model's flexible framework 
naturally extends to more complex astrophysical scenarios, demonstrating its broad potential for advancing our understanding of 
stellar activity and system architectures. Multi-planet systems represent a particularly compelling extension of our approach. 
The current framework can readily accommodate multiple transiting planets by incorporating additional transit signals into the design matrix, and therefore 
into the likelihood function. Each planet would contribute its own set of geometric parameters while sharing the same underlying 
stellar surface map and rotation properties. This extension would provide unprecedented opportunities to study stellar obliquity evolution 
across planetary systems and investigate potential dynamical interactions that might influence spot-crossing patterns across different orbital periods.

Eclipsing binary systems present another natural application for our model. The geometry of eclipsing binaries offers complementary constraints 
on stellar inclination and surface features, potentially breaking degeneracies that may persist in single-star planetary systems. 
The larger eclipse depths and longer eclipse durations in binary systems could provide enhanced sensitivity to spot-crossing events, 
enabling more detailed characterization of stellar surface features and their evolution.

The current paper is the first one in the series of works we plan on applying these techniques to extensive datasets from the 
Kepler and TESS missions. The forthcoming papers will demonstrate the method's practical utility for large-scale surveys and its 
potential for contributing to our understanding of stellar-planetary interactions in the era of precision photometry.

\section{Conclusions}
The \texttt{StarryStarryProcess} model presented in this work represents a significant advancement in our ability to characterize stellar surfaces using transit light curves. 
By combining spherical harmonic surface mapping, probabilistic spot modeling, and comprehensive transit analysis, we demonstrate that planetary transits provide unique constraints on 
stellar surface features that are not obtainable from rotational light curves alone.

Our key results include:
\begin{itemize}
    \item Transit light curves contain sufficient information to constrain the latitude distribution of starspots, particularly when multiple transit events are analyzed simultaneously. 
    This addresses a fundamental limitation of previous approaches that relied solely on rotational modulation.
    \item The model successfully recovers three-dimensional information about the star-planet system geometry, including stellar inclination and obliquity, using only photometric data. 
    \item For TOI-3884, we find compelling evidence for high-latitude spot concentrations ($\mu_\phi = 75.24^{+2.67}_{-4.11}$$^\circ$) and significant spin-orbit misalignment 
    ($\psi_\star = 80.37^{+49.6}_{-27.5}$$^\circ$).
    \item Our model accurately predicts transit light curve morphologies across multiple epochs, including subtle spot-crossing events that appear as small brightness increases during transit.
    \item The preliminary extension to evolving surfaces demonstrates the potential for tracking spot evolution over time, though additional development is needed for fully time-dependent modeling.
\end{itemize}
The inherent degeneracies in transit modeling -- particularly the ambiguity between northern and southern hemisphere spot locations and the inclination-obliquity degeneracy -- represent fundamental 
limitations that persist even with our approach. However, the consistency of the high-latitude spot configuration across different posterior samples indicates that our overall 
characterization of stellar activity patterns is robust.

More broadly, our work demonstrates that careful modeling of transit light curves can significantly improve both planetary characterization and stellar activity studies. 
By disentangling spot effects from planetary signals, we enable more accurate measurements of exoplanet properties while simultaneously extracting valuable information about stellar magnetic 
activity. This synergy will become increasingly important as missions like JWST attempt to characterize the atmospheres of Earth-sized planets around active host stars.

Future work should focus on incorporating multi-wavelength transit observations to better constrain spot temperatures and sizes, developing computationally efficient 
implementations that can handle higher-resolution surface maps, and refining time-dependent models to track spot evolution over longer baselines. 
The framework established here provides a foundation for these advances and opens new possibilities for studying stellar surfaces through 
the unique window provided by transiting exoplanets.

\section*{Acknowledgments}
S.S. thanks Adrian Price-Wheelan, Lionel Garcia, Charles Margossian, and the Data Group
and Gravitational Wave Astronomy Group at Flatiron Institute for
the valuable discussions. S.S. also acknowledges a graduate fellowship at the
Kavli Institute for Theoretical Physics, during which this research was
completed. S.S. thanks the LSSTC Data Science Fellowship Program, funded by
LSSTC, NSF Cybertraining Grant number 1829740, the Brinson Foundation, and the
Moore Foundation; her participation in the program has benefited this work. Finally, S.S. 
acknowledges support from award 644616 from the Simons Foundation.

\section*{Appendix A}
This study was carried out using the reproducibility software
\href{https://github.com/showyourwork/showyourwork}{\showyourwork}
\citep{Luger2021}, which leverages continuous integration to
programmatically download the data from
\href{https://zenodo.org/}{zenodo.org}, create the figures, and
compile the manuscript. Each figure caption contains two links: one
to the dataset stored on zenodo used in the corresponding figure with the icon \faDatabase,
and the other to the jupyter notebook used to make the figure with the icon \faGithub. The git
repository associated to this study is publicly available at
\url{https://github.com/ssagynbayeva/polka-dotted-stars-toi3884/tree/main}. 
The used chains data
are stored at \url{https://doi.org/10.5281/zenodo.16647404} and the notebooks are stored at 
\url{https://github.com/ssagynbayeva/polka-dotted-stars-toi3884/tree/main/src/tex/notebooks} to produce the figures.

\bibliography{bib}
\end{document}